\documentclass[twocolumn]{aastex631}

\accepted{October 15, 2022}

\submitjournal{ApJS}

\shorttitle{Time-Variable Jet Ejections from RW Aur A, RY Tau and DG Tau}
\shortauthors{Takami et al.}
\graphicspath{{./}{figures/}}
\usepackage{color}
\usepackage{soul,xcolor}
\setstcolor{red}

\begin{document}

\bibliographystyle{astron}

\title{Time-Variable Jet Ejections from RW Aur A, RY Tau and DG Tau
\footnote{
Based on observations obtained at the Gemini Observatory, the European Southern Observatory (under ESO programme 2100.C-5015), and W. M. Keck Observatory.
}}

\correspondingauthor{Michihiro Takami}
\email{hiro@asiaa.sinica.edu.tw}

\author[0000-0001-9248-7546]{Michihiro Takami}
%\author{Michihiro Takami}
\affil{Institute of Astronomy and Astrophysics, Academia Sinica, 
11F of Astronomy-Mathematics Building,
No.1, Sec. 4, Roosevelt Rd, Taipei 10617, Taiwan, R.O.C.}

\author[0000-0003-4243-2840]{Hans Moritz G\"unther}
%\author{Hans Moritz G\"unther}
\affil{MIT, Kavli Institute for Astrophysics and Space Research, 77 Massachusetts Avenue, Cambridge, MA 02139, USA}

\author[0000-0002-5094-2245]{P. Christian Schneider}
%\author{P. Christian Schneider}
\affil{Hamburger Sternwarte, Universit\"at Hamburg, Gojenbergsweg 112, D-21029, Hamburg, Germany}

\author[0000-0002-6881-0574]{Tracy L. Beck}
%\author{Tracy L. Beck}
\affil{The Space Telescope Science Institute, 3700 San Martin Dr., Baltimore, MD 21218, USA}

\author[0000-0003-3095-4772]{Jennifer L. Karr}
%\author{Jennifer L. Karr}
\affil{Institute of Astronomy and Astrophysics, Academia Sinica, 
11F of Astronomy-Mathematics Building,
No.1, Sec. 4, Roosevelt Rd, Taipei 10617, Taiwan, R.O.C.}

\author[0000-0001-9490-3582]{Youichi Ohyama}
%\author{Youichi Ohyama}
\affil{Institute of Astronomy and Astrophysics, Academia Sinica, 
11F of Astronomy-Mathematics Building,
No.1, Sec. 4, Roosevelt Rd, Taipei 10617, Taiwan, R.O.C.}

\author[0000-0003-1480-4643]{Roberto Galv\'an-Madrid}
%\author{Roberto Galv\'an-Madrid}
\affil{
Instituto de Radioastronom\'ia y Astrof\'isica, Universidad Nacional Aut\'onoma de M\'exico, Apdo. Postal 3-72 (Xangari), 58089 Morelia, Michoac\'an, Mexico
}

\author[0000-0002-6879-3030]{Taichi Uyama}
%\author{Taichi Uyama}
\affil{Infrared Processing and Analysis Center, California Institute of Technology, 1200 E. California Blvd., Pasadena, CA 91125, USA}
\affil{NASA Exoplanet Science Institute, Pasadena, CA 91125, USA}
\affil{National Astronomical Observatory of Japan, 2-21-1 Osawa, Mitaka, Tokyo 181-8588, Japan}

\author[0000-0003-3882-418X]{Marc White}
%\author{Marc White}
\affil{Research School of Astronomy and Astrophysics,
College of Physical \& Mathematical Sciences,
The Australian National University,
Mount Stromlo Observatory,
Cotter Rd,
Weston Creek ACT 2611,
Australia}

\author[0000-0001-5707-8448]{Konstantin Grankin}
\affil{Crimean Astrophysical Observatory, Russian Academy of Sciences, 298409 Nauchny, Crimea}

\author[0000-0002-2210-202X]{Deirdre Coffey}
%\author{Deirdre Coffey}
\affil{University College Dublin, School of Physics, Belfield, Dublin 4, Ireland}
\affil{Dublin Institute for Advanced Studies, School of Cosmic Physics, 31 Fitzwilliam Place, Dublin 2, Ireland}

\author[0000-0002-1624-6545]{Chun-Fan Liu}
%\author{Chun-Fan Liu}
\affil{Institute of Astronomy and Astrophysics, Academia Sinica, 
11F of Astronomy-Mathematics Building,
No.1, Sec. 4, Roosevelt Rd, Taipei 10617, Taiwan, R.O.C.}

\author[0000-0003-3500-2455]{Misato Fukagawa}
%\author{Misato Fukagawa}
\affil{National Astronomical Observatory of Japan, 2-21-1 Osawa, Mitaka, Tokyo 181-8588, Japan}

\author[0000-0001-7548-0190]{Nadine Manset}
\affil{Canada-France-Hawaii Telescope, 65-1238 Mamalahoa Hwy., Kamuela, HI 96743, USA}

\author[0000-0003-0262-272X]{Wen-Ping Chen}
%\author{Wen-Ping Chen}
\affil{Institute of Astronomy, National Central University, Taiwan 320, Taiwan}

\author[0000-0002-3273-0804]{Tae-Soo Pyo}
%\author{Tae-Soo Pyo}
\affil{Subaru Telescope, 650 North Aohoku Place, Hilo, HI 96720, USA}

\author[0000-0001-8385-9838]{Hsien Shang}
%\author{Hsien Shang}
\affil{Institute of Astronomy and Astrophysics, Academia Sinica, 
11F of Astronomy-Mathematics Building,
No.1, Sec. 4, Roosevelt Rd, Taipei 10617, Taiwan, R.O.C.}

\author[0000-0002-2110-1068]{Thomas P. Ray}
%\author{Thomas P. Ray}
\affil{Dublin Institute for Advanced Studies, School of Cosmic Physics, 31 Fitzwilliam Place, Dublin 2, Ireland}

\author[0000-0001-7076-0310]{Masaaki Otsuka}
%\author{Masaaki Otsuka}
\affil{Okayama Observatory, Kyoto University, Kamogata, Asakuchi, Okayama 719-0232, Japan}

\author[0000-0003-3602-0663]{Mei-Yin Chou}
%\author{Mei-Yin Chou}
\affil{Institute of Astronomy and Astrophysics, Academia Sinica, 
11F of Astronomy-Mathematics Building,
No.1, Sec. 4, Roosevelt Rd, Taipei 10617, Taiwan, R.O.C.}



\begin{abstract}
We present Gemini-NIFS, VLT-SINFONI and Keck-OSIRIS observations of near-infrared [\ion{Fe}{2}] emission associated with the well-studied jets from three active T Tauri stars; RW Aur A, RY Tau and DG Tau taken from 2012-2021. We primarily covered the redshifted jet from RW Aur A, and the blueshifted jets from RY Tau and DG Tau, to investigate long-term time variabilities potentially related to the activities of mass accretion and/or the stellar magnetic fields. All of these jets consist of several moving knots with tangential velocities of 70-240 km s$^{-1}$, ejected from the star with different velocities and at irregular time intervals. Via comparison with literature, we identify significant differences in tangential velocities for the DG Tau jet between 1985-2008 and 2008-2021. The sizes of the individual knots appear to increase with time, and in turn, their peak brightnesses in the 1.644-\micron~emission decreased up to a factor of $\sim$30 during the epochs of our observations. A variety of the decay timescales measured in the [\ion{Fe}{2}] 1.644 \micron~emission can be attributed to different pre-shock conditions if the moving knots are unresolved shocks. However, our data do not exclude the possibility that these knots are due to non-uniform density/temperature distributions with another heating mechanism, or in some cases due to stationary shocks without proper motions. Spatially resolved observations of these knots with significantly higher angular resolutions are necessary to better understand their physical nature.
\end{abstract}



%%%%%%%%%%%%%%%%%%%%%%%%%%%%%%%%%%%%%%%%
%%%%%%%%%%%%%%%%%%%%%%%%%%%%%%%%%%%%%%%%
%%%%%%%%%%%%%%%%%%%%%%%%%%%%%%%%%%%%%%%%
%%% 1. Introduction
%%%%%%%%%%%%%%%%%%%%%%%%%%%%%%%%%%%%%%%%
%%%%%%%%%%%%%%%%%%%%%%%%%%%%%%%%%%%%%%%%

\section{Introduction} \label{intro}

Young stellar objects (YSOs) of various masses and at various evolutionary stages are known to host collimated jets. Many of them, in particular those associated with Class I-II YSOs, are known to be associated with atomic and ionic emission lines at optical and infrared wavelengths. Jets from some young systems (Class 0-I) are associated with molecular line emission, in particular near-infrared H$_2$ and millimeter SiO/CO emission, while these lines are faint or absent in the more evolved phase (Class II or pre-main sequences). Furthermore, X-ray and/or centimeter continuum emission have been observed toward some jets. See \citet{Ray07,Frank14} for reviews of these observations.

Theoretical work over past decades has predicted that the jets play an essential role in protostellar evolution, removing excess angular momentum from accreting material and allowing mass accretion to occur \citep[e.g.,][]{Blandford82,Pudritz83,Shu00,Konigl00,Bai16}. This scenario has been supported by a statistical correlation between the observed mass ejection and accretion rates for many pre-main sequence stars \citep[e.g.,][]{Cabrit90,Hartigan95,Calvet97}, and observations of spinning motions in the jet \citep[e.g.,][]{Bacciotti02,Coffey04,Coffey07,Lee17}.
Understanding the jet driving mechanism and its detailed physical link with protostellar evolution are two of the most important issues of star formation theories.
%Furthermore, as YSOs are the nearest of a variety of astronomical jets with accretion disks (including X-ray binaries and active galactic nuclei), our understanding of YSOs may even be useful (or are fundamentally important) for understanding the physics of these systems in general.

Several theories have been proposed for the jet launching and driving, and their physical link with mass accretion. Popular magneto-centrifugal wind models have two main theories: (1) X-wind \citep{Shu00}, in which the jet launches from the inner edge of the disk ($r\ll0.1$ au); and (2) disk wind \citep{Konigl00}, in which the jet launching region covers a larger portion of the disk surface on a few au scale. Alternative mechanisms for jet driving include magnetic pressure \citep[e.g.,][]{Machida08} and reconnection of magnetic fields between the star and the disk \citep[reconnection wind, see e.g.,][for a review]{Bouvier14}. 
%Understanding the launching mechanism of the jets from YSOs is one of the biggest challenges for studies of star formation.
However, observational studies of the above theories have been hampered by the limited angular resolution of present telescopes (typically as good as $\sim 0\farcs1$, corresponding to $\sim$10 au in the nearest star forming regions) \cite[see][for a review]{Frank14}. 

Therefore, we have relied on observations of relatively extended parts ($\gtrsim$10 au from the star) of the jet to tackle the above issue. Some researchers have observed their spinning motions, and showed that these are consistent with the predictions of magneto-centrifugal wind models such as the X-wind and disk wind models \citep[e.g.,][]{Bacciotti02,Coffey04,Lee17}. \citet{Garufi19,Takami20} reported a possible time correlation between jet knot ejections from active pre-main sequence  stars and their potential signatures of mass accretion such as optical photometry and spectroscopy. 
The measurements by \citet{Takami20}  suggest that each of the jet knot ejections occurs within $\sim$100 days of an enhancement of mass accretion. Such a short delay time scale would be explained if the jet launching occurs within 0.1 au of the star.

Most of these studies are based on the observations of the jet for a single epoch or two, despite a timescale for their evolutions of millions of years \citep{Stahler05}. Although multi-epoch observations executable for human being are significantly shorter than the latter, long-term ($\gg$1 yr) monitoring observations of the jet are still useful for
investigating the stability of their physical conditions in order to study the evolutions of protostars and young stars. The time variabilities of the jet ejections from pre-main sequence stars are far less known than those of optical photometry and spectra, some of which are probably due to time variable mass accretion \citep[see][for reviews]{Bouvier07_PPV,Bouvier14}.
On the other hand, we can observe the jets from pre-main sequence stars ejected even hundred years ago \citep[e.g.,][]{Berdnikov17}. In this context, detailed studies of these jets are potentially useful for investigating the time variation of mass accretion and/or stellar activities at significantly longer timescales than those over the entire history of spectroscopic observations of pre-main sequence stars to date.

In this paper we present long-term monitoring data for the jets associated with three of the best-studied pre-main sequence stars: RW Aur A, RY Tau and DG Tau. We have monitored their jet ejections from 2012-2021 in [\ion{Fe}{2}] 1.644 \micron~emission, the brightest emission line in the near-infrared, using the technique of integral field spectroscopy with adaptive optics. The rest of the paper is organized as follows. In Section \ref{targets}, we summarize the understanding on these jets and host stars to date. In Section \ref{obs}, we describe the observations and data reduction. In Section \ref{results}, we present the results and analyze them, tentatively attributing the observed jet knots to `moving knots' as for many previous studies. In Section \ref{discussion} we summarize time variable jet ejections, including comparisons with literature, and discuss the physical nature of the knotty structures we observed. We give a summary and conclusions in Section \ref{summary}.

%%%%%%%%%%%%%%%%%%%%%%%%%%%%%%%%%%%%%%%%
%%%%%%%%%%%%%%%%%%%%%%%%%%%%%%%%%%%%%%%%
%%%%%%%%%%%%%%%%%%%%%%%%%%%%%%%%%%%%%%%%
%%% 2. Targets
%%%%%%%%%%%%%%%%%%%%%%%%%%%%%%%%%%%%%%%%
%%%%%%%%%%%%%%%%%%%%%%%%%%%%%%%%%%%%%%%%

\section{Targets} \label{targets}

In Table \ref{tbl:stars} we summarize the main properties of the target stars.
In Sections \ref{targets:rw}-\ref{targets:dg}, we summarize our understanding of individual target jets and host stars to date.

%, with the masses of 1--2 $M_\sun$ and the ages of 1-10 Myr at to distance $\sim$140 pc.

%%%%%%%%%%%%%%%%%%%%%%%%
%%% Table : Targets
\begin{table*}
\caption{Targets \label{tbl:stars}}
\begin{tabular}{lcccccc}
\tableline\tableline
%Semester\tablenotemark{a} & Run & Dates (YYYY-MM-DD)\\ \tableline
Star 	& Distance\tablenotemark{a}   	& Stellar Mass	& Age	& Mass Accretion Rate	& $v_{\mathrm{sys}}$\tablenotemark{b} & References\tablenotemark{c} \\
	& (pc) 					& ($M_\sun$)	& (Myr)	& (10$^{-7}$ $M_\sun$ yr$^{-1}$) & (km s$^{-1}$)
%Instrument  	&		&  (Selected) Sampling 	&		& ($\micron$)		& $R$				& (FoV)
\\ \tableline
RW Aur A	& 156$\pm$1\tablenotemark{d}	& 1.4$\pm$0.2	& 4$\pm$2 & 0.3	& 20	& 1\\
%RW Aur A	&  (A) 184$\pm$11 (B) \\
RY Tau	& 125$\pm$2 	& 2.0$\pm$0.3	&6$\pm$2		& 0.2-1.4	& 18	& 2,3\\
DG Tau	& 138$\pm$4	& $\sim$1 &	$\sim$1 	& 0.5-8 &	16 & 1,4,5,6\\
%2010B	& 1	&	2010-10-(16, 21)	\\
\tableline
\end{tabular} \\
\tablenotetext{a}{Based on the Gaia DR3 parallax measurements.}
\tablenotetext{b}{In the Heliocentric frame. Based on the references for Table \ref{tbl:vr_lit}.}
\tablenotetext{c}{(1) \citet{Dodin20}; (2)  \citet{Calvet04}; (3) \citet{Garufi19} ; (4) \citet{Muzerolle98_IR}  ; (5) \citet{Gullbring00}  ; (6) \citet{White04}  }
\tablenotetext{d}{The measurements for RW Aur A may suffer from the presence of a very close binary companion \citep[e.g.,][]{Petrov01a} or occultation by the dusty environment \citep[e.g.,][]{Schneider15,Dodin19}. We therefore adopt the distance to the binary companion RW Aur B, which is $\sim$1\farcs5 apart from RW Aur A. }
\end{table*}
%%%%%%%%%%%%%%%%%%%%%%%%

%%%%%%%%%%%%%%%%%%%%%%%%%%%%%%%%%%%%%%%%
%%% RW Aur A
%%%%%%%%%%%%%%%%%%%%%%%%%%%%%%%%%%%%%%%%

\subsection{RW Aur A} \label{targets:rw}

%RW Aur A is associated with a bipolar asymmetric jet, consisting of a brighter redshifted jet and a fainter blueshifted counterpart 
RW Aur A is associated with a brighter redshifted jet and a fainter blueshifted jet, extending over a few arcminute scale in opposite directions
\citep{Mundt98,Hirth94,Hirth97,Bacciotti96,Berdnikov17}. The asymmetry in jet emission is either due to different mass ejection rates between the redshifted and blueshifted jets \citep{Liu12}, or different physical conditions of surrounding gas on the two sides but with similar mass ejection rates \citep{Melnikov09}. 

The observed jets consist of 3-9 knots within 15\arcsec~ of the star.
These have been extensively observed at high-angular resolutions ($\sim$0\farcs1) at optical \citep[{[\ion{O}{1}] 6300/6363 \AA, [\ion{S}{2}] 6731/6716 \AA, [\ion{N}{2}] 6583 \AA} ---][]{Dougados00,Woitas02,Lopez03,Coffey08}
and near-infrared wavelengths \citep[{[}\ion{Fe}{2}{]} and H$_2$, mainly with the brightest lines at 1.644 \micron~and 2.122 \micron, respectively --- ][]{Pyo06,Beck08,Hartigan09,Takami20}
using \textit{The Space Telescope Imaging Spectrograph} (STIS) on  \textit{Hubble Space Telescope}, 
STIS2\footnote{The detector for HST-STIS}
and \textit{Optically Adaptive System for Imaging Spectroscopy} (OASIS) on \textit{the Canada-France-Hawaii Telescope} (CFHT) with the PUE'O adaptive optics (AO) system, 
\textit{The Infrared Camera and Spectrograph} (IRCS) on Subaru, 
\textit{Near-Infrared Integral Field Spectrometer} (NIFS) on the Gemini North telescope and
\textit{Near-InfRared Spectrograph} (NIRSpec) on the W. M. Keck II telescope.
%These observations have revealed a chain of knots in these jets within a few arcsecond of the star.
These observations measured radial velocities of the redshifted and blueshifted jets of 100 to 130 km s$^{-1}$ and --150 to --220 km s$^{-1}$, respectively, in the optical and near-infrared forbidden line emission described above.
\citet{Lopez03} measured proper motions of the jet knots of 0\farcs16--0\farcs26 yr$^{-1}$ at 1\arcsec--3\arcsec~from the star.
Some publications show their internal kinematics \citep{Woitas02,Pyo06,Coffey08,Hartigan09} including their spinning motions \citep{Coffey04,Coffey12}, while other publications derived detailed physical parameters such as electron densities, temperatures, and mass ejection rates \citep{Woitas02,Coffey08,Hartigan09}.

X-ray observations by \citet{Skinner14} showed that at least the redshifted jet appears to be associated with X-ray emission. The authors point out that the shock velocities inferred from optical and near-infrared observations are too low to explain this emission, suggesting the presence of another heating mechanism, e.g., via energy transfer from the star through the internal magnetic field, in the jet.

A number of optical and near-infrared spectroscopic observations have been made to understand magnetospheric accretion and wind activities close to the star \citep[e.g.,][]{Petrov01a,Alencar05,Takami16,Facchini16,Lisse22}.  The star appears to have been photometrically stable over many years \citep[e.g.,][]{Beck01,Grankin07}, however, it has shown peculiar photometric changes at a variety of wavelengths since 2010 \citep[$\Delta m_V$$\sim$3 mag., $\Delta m_K$$\sim$2 mag.; e.g.,][]{Rodriguez13,Rodriguez18,Schneider15,Petrov15,Shenavrin15,Bozhinova16,Lamzin17,Gunther18,Dodin19}.  Many of these authors attributed the photometric variations to occultations by dusty layers or blobs associated with the inner disk region or a wind. This explanation is corroborated by polarimetric observations by \citet{Dodin19}, which show a larger polarization in the faint state, indicating a larger contribution of scattered light to the observed brightnesses. See also \citet{Koutoulaki19} for the same interpretation with a near-infrared spectral variability.

\citet{Takami16,Takami20} observed different optical line profile variabilities between the bright and faint states, and discussed the possibility that the photometric variabilities are associated with mass accretion. Some line profiles show larger or more complicated time variations in the bright states, which can be attributed to occurrence of magneto-hydrodynamical (MHD) instabilities of accretion flows at high mass accretion rates \citep{Romanova08,Kurosawa13}. \citet{Takami20} reported a possible correlation between these variabilities and jet knot ejections. Remarkable optical line profile changes have also been observed by \citet{Petrov01a,Petrov01b,Petrov07,Chou13}.

While RW Aur A is associated with a resolved companion 1\farcs5 away \citep[RW Aur B, e.g.,][]{Joy44,Reipurth93,White01,Bisikalo12}, a few spectroscopic studies suggest that RW Aur A itself is a spectroscopic binary \citep[e.g.,][]{Gahm99,Petrov01a}.

%%%%%%%%%%%%%%%%%%%%%%%%%%%%%%%%%%%%%%%%
%%% RY Tau
%%%%%%%%%%%%%%%%%%%%%%%%%%%%%%%%%%%%%%%%

\subsection{RY Tau} \label{targets:ry}

As with RW Aur A, RY Tau is associated with a bipolar jet. \citet{St-Onge08} showed that the blueshifted jet extends out to at least 31\arcsec~from the star. 
The redshifted jet is much fainter at this angular scale, probably due to obscuration by a dusty circumstellar disk \cite[e.g.,][]{Isella10} and/or a remnant envelope \citep{Takami13,Garufi19} as is the case for many other low-mass pre-main sequence stars \citep[see, e.g.,][for a review]{Eisloffel00_PPIV}. \citet{St-Onge08} alternatively identified two bow shocks associated with the redshifted jet, 2\farcm8 and 3\farcm1 away from the star.

Spatially resolved imaging observations of the blueshifted jet have been made by several groups in H$\alpha$ 6563 \AA~emission \citep{St-Onge08,Uyama22}, ultraviolet \ion{C}{4} emission \citep[1548/1551 \AA;][]{Skinner18}, and low-excitation forbidden emission lines at optical \citep[{[\ion{O}{1}] 6300 \AA;}][]{Agra09} and near-infrared wavelengths \citep[{[\ion{Fe}{2}] 1.644 \micron;}][]{Coffey15,Uyama22}. \citet{Garufi19} presented observations of all of these lines as well as near-infrared [\ion{S}{2}] 1.029-1.037 \micron  and \ion{He}{1} 1.083 \micron lines. 
The presence of high excitation lines such as H$\alpha$, \ion{He}{1} and \ion{C}{4} lines may be due to shocks that are more energetic than those of  jets associated with many other pre-main sequence stars \citep{Eisloffel00_PPIV,Hartigan00_PPIV}. \citet{Skinner11} reported probable detection of X-ray emission in the jet. 

Most of these observations at optical and near-infrared wavelengths have been made at high angular resolutions comparable to or better than 0\farcs4, with the best resolutions of 0\farcs03--0\farcs05 \citep{Garufi19,Uyama22} using 
\textit{the Wide Field and Planetary Camera 2} (WFPC2) and STIS on the HST,
CFHT-OASIS,
GEMINI-NIFS,
\textit{Spectro-Polarimetric High-contrast Exoplanet REsearch} SPHERE on \textit{the Very Large Telescope} and
\textit{The Visible Aperture Masking Polarimetric Interferometer for Resolving Exoplanetary Signatures} (VAMPIRES) on Subaru with the \textit{Subaru Coronagraphic Extreme AO} (SCExAO).
%These observations have revealed a chain of jet knots.

In the images obtained by \citet{St-Onge08}, the blueshifted jet consists of several knots at a $\sim$30\arcsec~scale, although the jet structures are not clear close to the star because of the bright stellar continuum. \citet{Garufi19,Uyama22} conducted integral field spectroscopy, which minimizes the stellar continuum emission, and revealed the presence of a few jet knots within $\sim$1\arcsec~of the star.

%%% but the structures are more extended downstream \citep[$\sim$5\arcsec~from the star;][]{Garufi19}.
%It is likely that a redshifted jet also exists, but it has not been observed as for many other low-mass pre-main sequence stars \citep[see, e.g.,][for a review]{Eisloffel00_PPIV} probably due to obscuration by a dusty circumstellar disk \cite[e.g.,][]{Isella10} and/or a remnant envelope \citep{Takami13,Garufi19}. 
\citet{Skinner18} revealed the presence of a faint redshifted jet within 0\farcs8 of the star, which has not been identified by any of the above high-resolution observations at optical and near-infrared wavelengths.
None of these high-resolution observations have shown the presence of a close companion within 1\arcsec~of the star.

\citet{Agra09,Coffey15} measured a velocity of the blueshifted jet of --60 to --90 km s$^{-1}$ in lowly excited forbidden lines ([\ion{O}{1}], [\ion{Fe}{2}]). This contrasts with the observations by \citet{Skinner18}, who measured --136$\pm$10 km  s$^{-1}$ in the \ion{C}{4} lines. This discrepancy indicates the presence of multiple velocity components traced by emission lines at different excitation conditions.

This star has long been subject to extensive photometric monitoring, exhibiting peculiar photometric variations with a $V$-band amplitude $\Delta m_V$ of $\sim$2.5 mag. \citep[e.g.,][]{Zajtseva85,Herbst84,Bouvier93,Herbst94,Petrov99,Grankin07,Garufi19}. As for RW Aur A, many of these authors attributed the photometric variabilities to dust occultations, while \citet{Garufi19} showed a possible correlation between photometric variability and jet knot ejections, suggesting that it is related to time variable mass accretion. This star is also known to exhibit remarkable variabilities in optical line profiles \citep{Petrov99,Petrov19,Petrov21,Chou13}. The observed timescales of line profile changes ranges from a few days to years. 
 %
%\cite{Petrov99} and \cite{Grankin07} monitored RY~Tau's optical spectrum/flux and \cite{Chou13} investigated mass accretion signatures such as H$\alpha$, which presented short-term variabilities ($\lesssim1$~yr).

%%%%%%%%%%%%%%%%%%%%%%%%%%%%%%%%%%%%%%%%
%%% DG Tau
%%%%%%%%%%%%%%%%%%%%%%%%%%%%%%%%%%%%%%%%

\subsection{DG Tau} \label{targets:dg}
The jet from DG Tau has been extensively observed for at least 40 years. The blueshifted jet was first identified as a single knot 8\arcsec away from the star \citep[][]{Mundt83}. Later on, a number of spectro-imaging observations at subarcsecond resolutions have been conducted in the optical \citep[H$\alpha$, {[\ion{O}{1}]} 6300/6363 \AA, {[\ion{S}{2}]} 6731/6716 \AA, {[\ion{N}{2}]} 6548/6583 \AA --- ][]{Kepner93,Lavalley00,Dougados00,Coffey07,LiuC16} and near-infrared \citep[{[\ion{Fe}{2}] 1.644 \micron;}][]{Pyo03,Agra11,White14a} using
CFHT-PUE'O+OASIS/STIS2,
HST-STIS,
Subaru-IRCS,
\textit{the Spectrograph for Integral Field Observations in the Near Infrared} (SINFONI) on VLT, and
Gemini-NIFS.
In contrast to the similar observations for RW Aur A and RY Tau, these observations with $\sim$0\farcs1 resolutions have revealed internal structures in the jet within 1\arcsec, which look similar to bow shocks and bubbles \citep{Bacciotti00,Agra11,White14a}. These observations have simultaneously revealed an ``onion-like" kinematic structure, with the fastest flow at the jet axis surrounded by slower flow components.
The presence of the optical [\ion{Ne}{3}] \citep[3869 \AA;][]{LiuC16} and infrared \ion{He}{1} \citep[1.083 \micron;][]{Takami02b} lines indicates a more energetic nature of the jet than those associated with many other pre-main sequence stars. The measured radial velocities of the jet in the above literature range from --120 to --350 km s$^{-1}$ (see Section \ref{discussion:moving:v} for details).

X-ray emission in the jet was observed by \citet{Gudel05,Gudel08,Schneider08}. \citet{Gudel11,White14a} reported the presence of a stationary shock component at $\sim$0\farcs2 from the star based on multi-epoch studies. Other detailed studies of the physical conditions of the jet include \citet{Coffey08,White14b,White16}. While many of the studies in the optical and near-infrared show only the blueshifted jet, \citet{Agra11,White14b,White14a} observed a faint redshifted bubble-like structure about 1\arcsec~away from the star in the opposite direction from the blueshifted jet.

\citet{Grankin07} measured a $V$-band photometric variability of the star of $\Delta m_V$$\sim$2.5 mag. over 20 years. \citet{Chou13} observed relatively stable optical line profiles over a few months in 2010, but found some differences from the literature based on the observations from 1983--1996, perhaps due to long-term variabilities on 10-30 year scales. As for RY Tau, the above high angular resolution observations at optical and near-infrared wavelengths have not shown the presence of a close binary companion.

%%%%%%%%%%%%%%%%%%%%%%%%%%%%%%%%%%%%%%%%
%%%%%%%%%%%%%%%%%%%%%%%%%%%%%%%%%%%%%%%%
%%%%%%%%%%%%%%%%%%%%%%%%%%%%%%%%%%%%%%%%
%%% 3. Observations and Data Reduction
%%%%%%%%%%%%%%%%%%%%%%%%%%%%%%%%%%%%%%%%
%%%%%%%%%%%%%%%%%%%%%%%%%%%%%%%%%%%%%%%%

\section{Observations and Data Reduction} \label{obs}

The observations were made using Gemini-NIFS, VLT-SINFONI and \textit{the OH-Suppressing Infrared Imaging Spectrograph} (OSIRIS) on Keck with adaptive optics. Table \ref{tbl:instruments} summarize the instrument specifications with the selected gratings and integral field units. The NIFS and SINFONI spectra cover several [\ion{Fe}{2}] lines including those at 1.644, 1.534, 1.600, 1.664, 1.712 and 1.745 $\micron$. The OSIRIS spectra cover a few major [\ion{Fe}{2}] lines among them at 1.59-1.67 \micron. The spectral resolutions of 3000-5300 are not sufficient for resolving internal kinematics of the target jets in many cases, but optimum for obtaining the images of emission lines with high signal-to-noise (Section \ref{results:v}). 

%%%%%%%%%%%%%%%%%%%%%%%%
%%% Table : Instruments
\begin{table*}
\caption{Instruments \label{tbl:instruments}}
\begin{tabular}{lcccccc}
\tableline\tableline
%Semester\tablenotemark{a} & Run & Dates (YYYY-MM-DD)\\ \tableline
Telescope  \& 	& Operation & IFU \&				& Grating	& Spectral Coverage	& Spectral Resolution	& Field of View \\
Instrument  	&		&  (Selected) Sampling 	&		& ($\micron$)		& $R$				& (FoV)
\\ \tableline
Gemini-NIFS	& Queue	& Slit Slicer, 0\farcs1$\times$0\farcs04 	& $H$ 	& 1.49-1.80 	& 5300		& 3\farcs0$\times$3\farcs0 \\
VLT-SINFONI	& Queue	& Slit Slicer, 0\farcs1$\times$0\farcs05	& $H$	& 1.45-1.85	& 3000		& 3\farcs6$\times$3\farcs3 \\
Keck-OSIRIS	& Classical & Lenslet, 0\farcs05				& Hn3	& 1.59-1.67	& 3800		& 3.2\arcsec$\times$2.4\arcsec\\
%2010B	& 1	&	2010-10-(16, 21)	\\
\tableline
\end{tabular} \\
\end{table*}
%%%%%%%%%%%%%%%%%%%%%%%%

Table \ref{tbl:log} shows the log of the observations for the three target stars. Many of the spectra were obtained using NIFS in photometric conditions with an angular resolution of typically 0\farcs10-0\farcs15. The data from 2012--2013 were obtained using an occulting disk at the focal plane with a 0\farcs2 diameter. The other data were obtained without an occulting mask, with short exposures to avoid saturation of the stellar continuum.

%%%%%%%%%%%%%%%%%%%%%%%%
%%% Table : Log
%\begin{rotatetable}
\begin{deluxetable*}{lcllcrrclcc}
%\tablenum{1}
\tablecaption{Log of the observations\label{tbl:log}}
%{\scriptsize
\tabletypesize{\scriptsize}
\tablewidth{0pt}
\tablehead{
\colhead{Star} &
\colhead{Date} &
\colhead{Instrument} &
\colhead{Run ID ( PI)} &
\colhead{Photometrically} &
\colhead{$t_{exp}$} &
\colhead{$n_{exp}$} &
\colhead{Core FWHM} &
\colhead{$f_{core}$\tablenotemark{b}} &
\multicolumn{2}{c}{Range of Integration \tablenotemark{c}}
%\multicolumn{2}{c}{Range for Integration}
%
\\
&&&&
\colhead{Accurate\tablenotemark{a}}& 
\colhead{(s)} && 
\colhead{(arcsec)} && 
\colhead{$v$ (km s$^{-1}$)} &
\colhead{$Y$\tablenotemark{d} (arcsec)}
}
%\decimalcolnumbers
\startdata
RW Aur A	& 2012 Oct 20 	& NIFS\tablenotemark{e}	& GN-2012B-Q-99 (Beck) 	& $\circ$ & 600	& 9		& 0.16	& 0.61	& 30 to 180 & --0.15 to +0.15\\
		& 2014 Feb 28	& NIFS		& GN-2014A-Q-29 (G\"unther)	& $\circ$	& 60		& 12		& 0.15	& 0.48	& 30 to 180 & --0.15 to +0.15\\
		& 2014 Dec 29	& NIFS		& GN-2014B-Q-18 (G\"unther)	& $\circ$	& 84		& 20		& 0.15	& 0.45	& 30 to 180 & --0.15 to +0.15\\
		& 2017 Feb 15	& NIFS		& GN-2017A-FT-1 (Takami)	& $\circ$	& 55		& 36		& 0.15	& 0.49	& 30 to 180 & --0.15 to +0.15\\
		& 2017 Dec 08	& SINFONI	& 2100.C-5015(A) (Takami) 	& 		& 4		& 140	& 0.10	& 0.24	&--20 to 230 & --0.15 to +0.15\\
		& 2017 Dec 11	& SINFONI	& 2100.C-5015(A) (Takami) 	& 		& 4		& 140	& 0.10	& 0.26	& --20 to 230 & --0.15 to +0.15\\
		& 2018 Aug 21	& NIFS		& GN-2018B-Q-141 (Takami) 	& 		& 55		& 17		& 0.14	& 0.46 	& 30 to 180 & --0.15 to +0.15\\
		& 2018 Aug 31	& NIFS		& GN-2018B-Q-141 (Takami)	& $\circ$	& 55		& 3		& 0.12	& 0.51	& 30 to 180 & --0.15 to +0.15\\
		& 2018 Sep 16	& NIFS		& GN-2018B-Q-141 (Takami)	& 		& 55		& 19		& 0.12	& 0.50	& 30 to 180 & --0.15 to +0.15\\
		& 2019 Oct 07	& NIFS		& GN-2019B-Q-132 (Takami)	& 		& 55		& 36		& 0.16	& 0.46	& 30 to 180 & --0.15 to +0.15\\
		& 2021 Feb 03	& OSIRIS		& S21A0039N/S364 (Takami)	& $\circ$ 	& 24		& 30		& 0.05-0.10\tablenotemark{f}	& $\sim$0.3\tablenotemark{f}	& 10 to 190 & --0.15 to +0.15 \\
RY Tau	& 2012 Oct 27 	& NIFS\tablenotemark{e}		& GN-2012B-Q-99 (Beck)	& $\circ$	& 600	& 10		& 0.14	& 0.52		& --170 to --40 & --0.25 to +0.25\\
		& 2014 Feb 28	& NIFS		& GN-2014A-Q-29 (G\"unther)	& $\circ$	& 15		& 30		& 0.18	& 0.35	& --170 to --40 & --0.25 to +0.25\\
		& 2014 Dec 29	& NIFS		& GN-2014B-Q-18 (G\"unther)	& $\circ$	& 15		& 54		& 0.18	& 0.46	& --170 to --40 & --0.25 to +0.25\\
		& 2017 Feb 18	& NIFS		& GN-2017A-FT-1 (Takami)	& $\circ$	& 15		& 105	& 0.12	& 0.46	& --170 to --40 & --0.25 to +0.25\\
		& 2018 Aug 17	& NIFS		& GN-2018B-Q-141 (Takami) 	& $\circ$	& 15		& 99		& 0.15	& 0.37 	& --170 to --40 & --0.25 to +0.25\\
		& 2019 Oct 23	& NIFS		& GN-2019B-Q-132 (Takami) 	& 		& 15		& 108	& 0.13	& 0.53	& --170 to --40 & --0.25 to +0.25\tablenotemark{g}\\
		& 2021 Feb 03	& OSIRIS		& S21A0039N/S364 (Takami) 	& 		& 19		& 46		& 0.05-0.10\tablenotemark{f}	& $\sim$0.4\tablenotemark{f}	& --160 to --30 & --0.2 to +0.2\tablenotemark{g}\\
DG Tau	& 2013 Feb 09 	& NIFS\tablenotemark{e}	& GN-2012B-Q-32 (McGregor)	& $\circ$	& 600	& 6		& 0.16	& 0.61	& --240 to --70 & --0.3 to +0.3 \\
		& 2014 Feb 28	& NIFS		& GN-2014A-Q-29 (G\"unther)	& $\circ$	& 45		& 12		& 0.12	& 0.50	& --240 to --70 & --0.3 to +0.3\\
		& 2014 Dec 29	& NIFS		& GN-2014B-Q-18 (G\"unther)	& $\circ$	& 45		& 27		& 0.14	& 0.37	& --240 to --70 & --0.3 to +0.3\\
		& 2017 Feb 17	& NIFS		& GN-2017A-FT-1 (Takami)	& $\circ$	& 25		& 72		& 0.11	& 0.44	& --240 to --70 & --0.3 to +0.3\\
		& 2017 Dec 22	& SINFONI	& 2100.C-5015(A) (Takami) 	& 		& 4		& 140	& 0.11	& 0.18	& --240 to --40 & --0.3 to +0.3\\
		& 2017 Dec 25	& SINFONI	& 2100.C-5015(A) (Takami) 	& 		& 4		& 140	& 0.11	& 0.20	& --240 to --40 & --0.3 to +0.3\\
		& 2018 Nov 27	& NIFS		& GN-2018B-Q-141 (Takami)	& $\circ$	& 25		& 72		& 0.12	& 0.48 	& --240 to --70 & --0.3 to +0.3\\
		& 2019 Oct 17	& NIFS		& GN-2019B-Q-132 (Takami) 	& $\circ$	& 25		& 84		& 0.16	& 0.37	& --240 to --70 & --0.3 to +0.3\\
\enddata
%}
\tablenotetext{a}{With an absolute photometric accuracy of $<$10 \%.}
\tablenotetext{b}{Fractional flux of the core of the point-spread function. (See text.)}
\tablenotetext{c}{For analysis in Section \ref{results}. 
%{\bf We carefully adjusted each range to cover most of the line emission but also maximize the signal-to-noise.}
}
\tablenotetext{d}{Across the jet axis.}
\tablenotetext{e}{The central star was masked using an occulting mask with a 0\farcs2 diameter.}
\tablenotetext{f}{Less accurate due to non-linear response at the brightest pixels. We have also used data with short exposures to derive these values.}
\tablenotetext{g}{Offset by 0\farcs15 for knot `E' for Figure \ref{fig:rw} in order to cover the emission (Section \ref{results:y}).}
\end{deluxetable*}
%\end{rotatetable}
%%%%%%%%%%%%%%%%%%%%%%%%

For many of the observations the star was placed near the edge of the field of view (FoV) to maximize coverage of the jet over a large spatial area. RW Aur A is associated with a bipolar jet, and we covered the redshifted jet (i.e., the brighter jet). For RY Tau and DG Tau, we covered the blueshifted jet as the redshifted counterpart is faint or absent due to obscuration by a circumstellar disk (Section \ref{targets}).

Data were reduced using the pipelines provided by the observatories, and software we developed using PyRAF, numpy, scipy and astropy on python. For the NIFS data, we used the Gemini IRAF package for sky subtraction, flat-fielding, the first stage of bad pixel removals, 2 to 3 dimensional transformation of the spectral data, and wavelength calibration. We then used our own software for stacking the data cubes for each date, telluric correction, flux calibration, extraction of the cube for the target emission line, additional removal of bad pixels, and continuum subtraction. We have also corrected for a flux loss with the halo of the point-spread function (PSF), as the jet structures we are interested in are significantly smaller than the halo of the PSF, which extends over a $>$0\farcs5 scale. We used the identical processes for the SINFONI and OSIRIS data, except that data stacking was made using the observatory pipeline.

Some queue observations were split into different nights within a month timescale. These data were stacked for individual autumn-winter periods. For each data set, we identify possible changes in the brightness of the knots within a month, perhaps due to changes in shock conditions on a month scale (Section \ref{discussion:moving:shocks}). 
Detailed analysis of the variation of jet emission on this time scale is beyond the scope of this study. We assume that these possible changes of jet emission are independent of the variabilities of mass accretion and/or the optical emission lines close to the star, which also show such short-term variabilities (Section \ref{discussion:moving:acc}). This is because these activities cannot physically interact with the jet knots once they move away from the star.

%As shown in Section \ref{results}, we do not expect significant time variation of jet emission over a month.

For the SINFONI and OSIRIS data, which have spectral resolutions lower than NIFS, we have made additional correction of slight errors in wavelength calibration using the telluric and photospheric absorption lines. As a result, we are confident of an accuracy for the measured absolute velocities at about a $\pm$10 km s$^{-1}$ level for the NIFS data, and $\pm$20 km s$^{-1}$ for the SINFONI and OSIRIS data.

For RW Aur A we found a marginal error ($\sim$1$\degr$) in the actual image position angle (PA) from those set for the NIFS, SINFONI and OSIRIS observations. This was corrected by measuring the PA toward the binary companion RW Aur B ($d \sim$ 1\farcs5). See \citet{Takami20} for details. As RY Tau and DG Tau are single stars, we regard the above error as a typical uncertainty of the jet PAs. This error probably explains slightly different jet PA for different epochs, which hampers reliable analysis. For those with relatively large PA offsets, we visually inspected the offsets and adjusted it (see Figures \ref{fig:rw}-\ref{fig:dg}) for our positional analysis along the jet axis in Section \ref{results}.

For this paper, we limit our analysis of the spatial distribution and kinematics to the [\ion{Fe}{2}] 1.644 \micron~ emission, and of the intensity ratios at the peaks of knots to the  [\ion{Fe}{2}] 1.644, 1.533 and 1.600 \micron~ emission due to limited signal-to-noise.
Some velocity-integrated maps for [\ion{Fe}{2}] 1.644 \micron~have already been published by \citet{Takami20,Uyama22}. \citet{Takami20} used all the data for RW Aur A but for the latest epoch (2021 February 3) and performed comparisons with jet knot ejections and optical photometry and spectroscopy to investigate a physical link between mass accretion and ejection. \citet{Uyama22} used the velocity-integrated maps for RY Tau for the latest two epochs (2019 October 23, 2021 February 03) and performed comparisons with jet knots seen in the H$\alpha$ emission. 

%%%%%%%%%%%%%%%%%%%%%%%%%%%%%%%%%%%%%%%%
%%%%%%%%%%%%%%%%%%%%%%%%%%%%%%%%%%%%%%%%
%%%%%%%%%%%%%%%%%%%%%%%%%%%%%%%%%%%%%%%%
%%% 4. Results
%%%%%%%%%%%%%%%%%%%%%%%%%%%%%%%%%%%%%%%%
%%%%%%%%%%%%%%%%%%%%%%%%%%%%%%%%%%%%%%%%
%%%%%%%%%%%%%%%%%%%%%%%%%%%%%%%%%%%%%%%%

\section{Results} \label{results}

\begin{figure*}[ht!]
\plotone{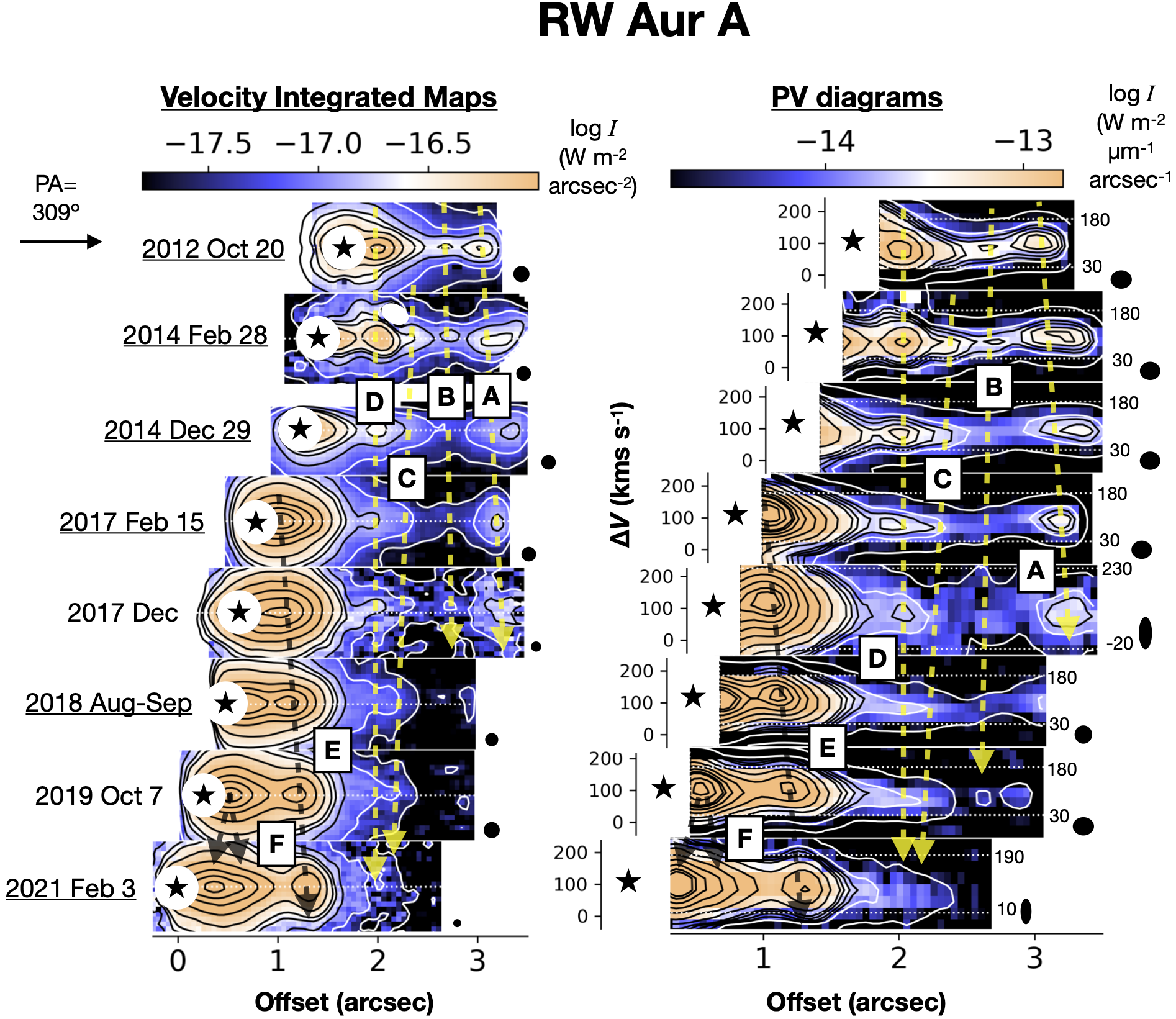}
\caption{The velocity-integrated maps (left) and position-velocity (PV) diagrams (right) of the [\ion{Fe}{2}] 1.644 \micron~emission associated with the redshifted jet from RW Aur A.
The top to bottom panels show the maps and diagrams for the eight epochs (see Table \ref{tbl:log} for details).
The data for the underlined dates were obtained with a photometric accuracy within 10 \%.
The velocities for the PV diagrams are in terms of the systemic velocity of the star.
The region within 0\farcs2 of the star is masked as imperfect subtraction of the bright continuum makes the data unreliable.
The contour levels
%(log $I$ [W m$^{-2}$ arcsec$^{-2}$]=-17.6/-17.2/-16.88/-16.66/-16.54/-16.3/-16/-15.7/-15.4/-15.1 for the velocity-integrated maps; log $I$ [W m$^{-2}$ \micron$^{-1}$ arcsec$^{-1}$]=-14.5/-13.9/-13.76/-13.65/-13.4/-12.9/-12.65/-12.45/-12.30/-12.2/-12.15/-12.02/-11.93 for the PV diagrams)
are arbitrarily chosen to show approximate locations of the jet knots.
The positions of the maps and the diagrams are offset from those at the bottom (i.e., those for the latest epoch) by 0\farcs2 yr$^{-1}$ to be able to easily identify the moving jet knots observed at different epochs.
The dashed arrows in the vertical direction indicate the identified moving knots.
We select either yellow or black for each arrow for visibility.
The white horizontal dotted lines in each PV diagram show the range used for making the velocity-integrated maps.
Their actual numbers are shown at the right side of these lines.
The black circles and ellipses at the right side of the individual panels show the angular and velocity resolution.
\label{fig:rw}}
\end{figure*}

\begin{figure*}[ht!]
\plotone{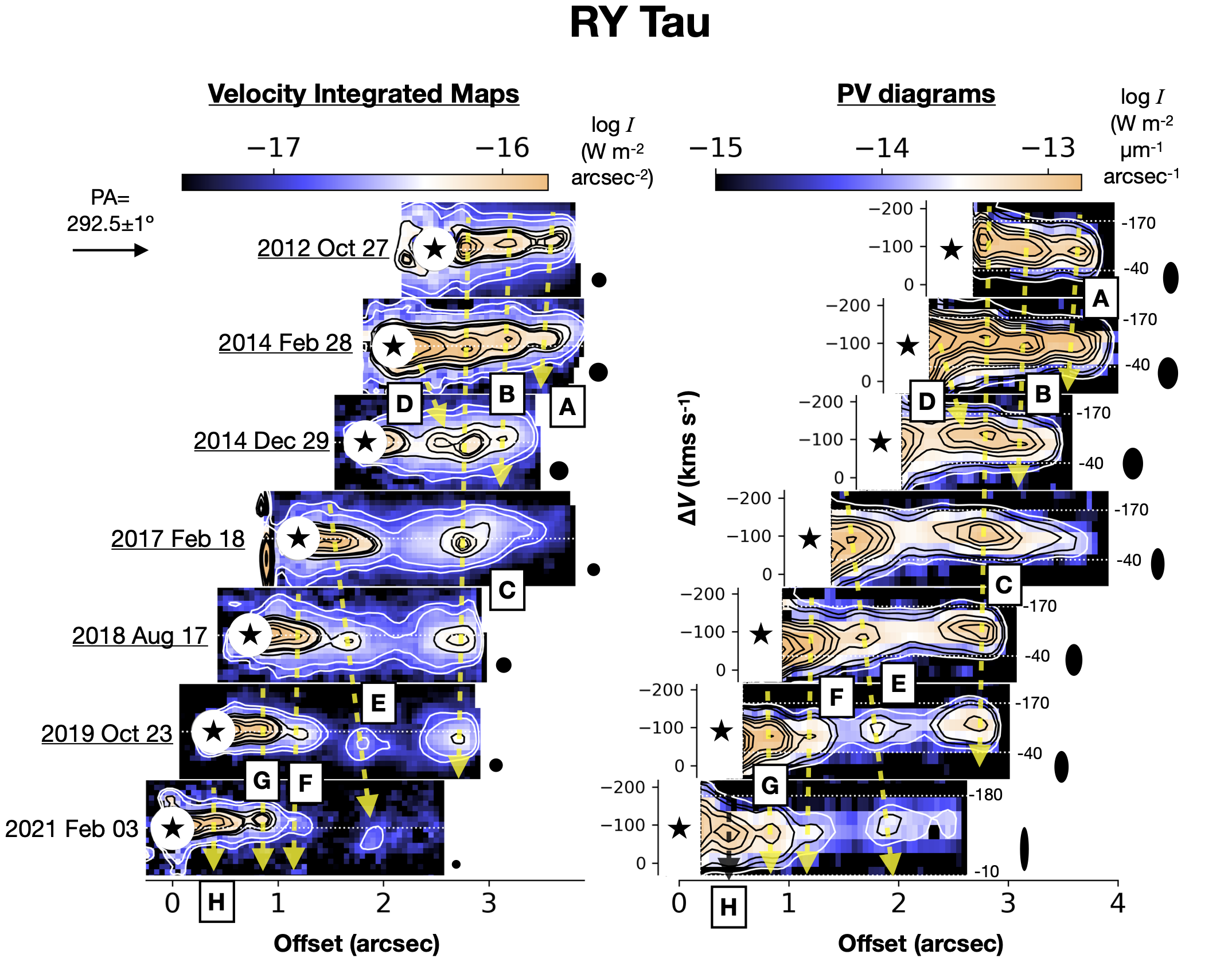}
\caption{Same as  Figure \ref{fig:rw} but for the blueshifted jet from RY Tau for seven epochs. The positions of the maps and the diagrams are offset from those at the bottom (i.e., those for the latest epoch) by 0\farcs3 yr$^{-1}$ to be able to easily identify the moving jet knots observed at different epochs.
\label{fig:ry}}
\end{figure*}

\begin{figure*}[ht!]
\plotone{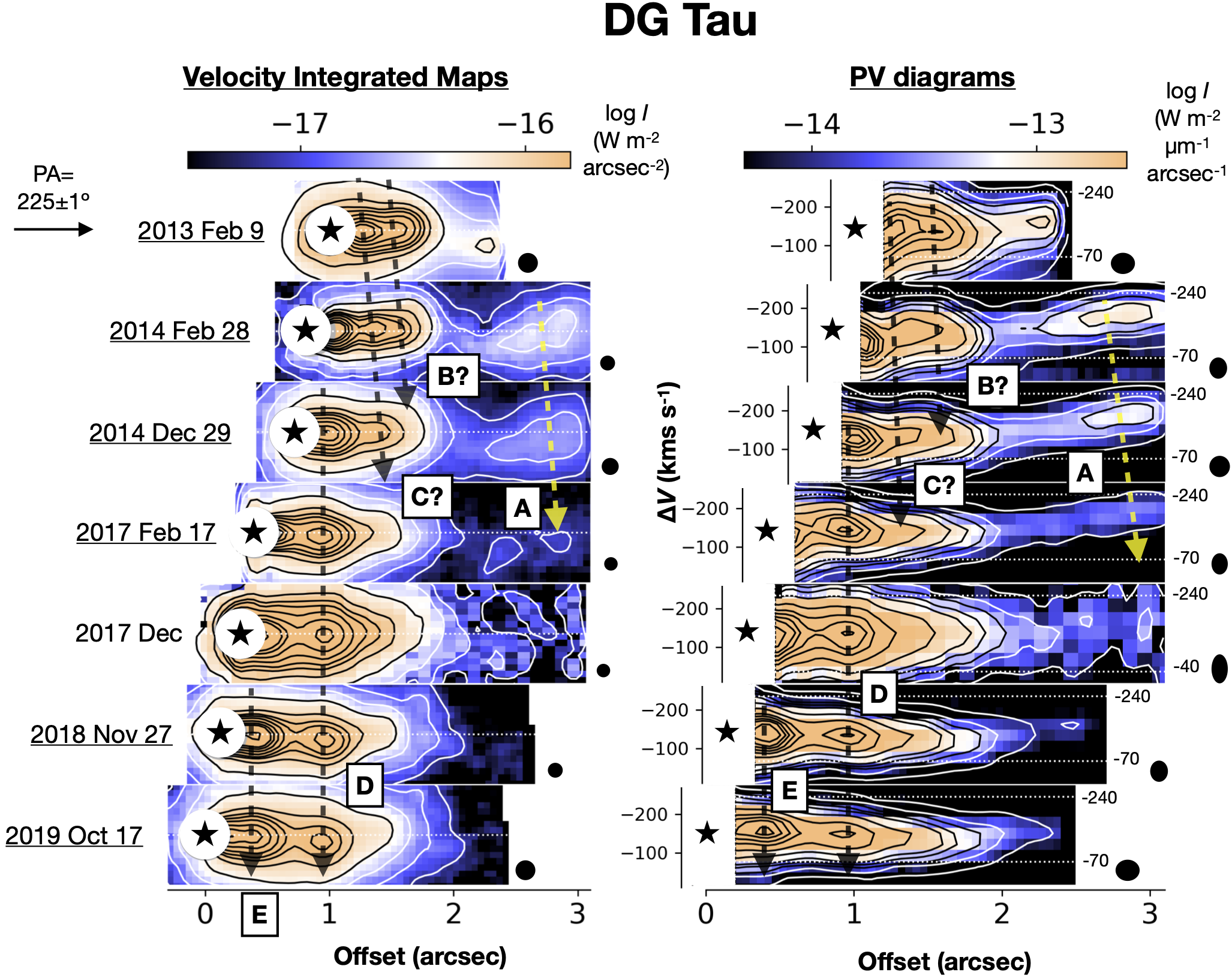}
\caption{Same as  Figures \ref{fig:rw} and \ref{fig:ry} but for the blueshifted jet from DG Tau for seven epochs. The positions of the maps and the diagrams are offset from those at the bottom (i.e., those for the latest epoch) by 0\farcs15 yr$^{-1}$ to be able to easily identify the moving jet knots observed at different epochs.
The furtherest bright blob seen in  the 2013 data appears to be the tail of the structure A but appearing a knot due to imperfect performance of the plotting tool near the edge of the FoV.
\label{fig:dg}}
\end{figure*}

Figures \ref{fig:rw}-\ref{fig:dg} show the velocity-integrated maps and the position-velocity (PV) diagrams of the [\ion{Fe}{2}] 1.644 \micron~emission for the redshifted jet from RW Aur A, the blueshifted jet from RY Tau, and the blueshifted jet from DG Tau, respectively. In each figure, we place the maps and diagrams in chronological order from top to bottom. In Table \ref{tbl:log} we tabulate the ranges for velocity and spatial integrations, which we carefully adjusted to cover most of the line emission but also maximize the signal-to-noise. The [\ion{Fe}{2}] 1.533 and 1.600 \micron~emission, which we will analyze later for intensity ratios, have spatial distributions very similar to the 1.644 \micron~emission but with low signal-to-noise due to their faint nature.

%%%%%%%%%%%%%%%%%%%%%%%%
%%% Figure: integrated maps & PV diagrams (RW Aur A)

%%%%%%%%%%%%%%%%%%%%%%%%
%%% Figure: integrated maps & PV diagrams (RY Tau)

%%%%%%%%%%%%%%%%%%%%%%%%
%%% Figure: integrated maps & PV diagrams (DG Tau)
%%%%%%%%%%%%%%%%%%%%%%%%

In these figures we identify chains of knotty structures as for previous spectro-imaging at ultraviolet to near-infrared wavelengths (see Section \ref{targets}). 
In this section we analyze these knots, $tentatively$ attributing them to moving knots as for several studies 
(see Table \ref{tbl:vt_lit} in Section \ref{discussion:moving:v}).
In Section \ref{results:x} we identify these knots,  then analyze their proper motions and the time intervals of the ejections.
In Section \ref{results:y} we briefly summarize their spatial extension across the jet axis.
In Section \ref{results:I} we statistically analyze their peak intensities, intensity ratios and inferred electron densities.
In Section \ref{results:v} we summarize the observed radial velocities and perform comparisons with tangential velocities inferred from Section \ref{results:x}.

Some knots may be alternatively attributed to `stationary shocks' without proper motions rather than the moving knots. We will discuss this issue in Section \ref{discussion:stationary}.

%%%%%%%%%%%%%%%%%%%%%%%%%%%%%%%%%%%%%%%%
%%%%%%%%%%%%%%%%%%%%%%%%%%%%%%%%%%%%%%%%
%%% 3.1, spatial distributions along the jet axis
%%%%%%%%%%%%%%%%%%%%%%%%%%%%%%%%%%%%%%%%
%%%%%%%%%%%%%%%%%%%%%%%%%%%%%%%%%%%%%%%%

\subsection{Identification of Moving Knots} \label{results:x}

To easily identify the moving jet knots observed at different epochs, we apply spatial offsets along the jet axis to individual panels corresponding to 0\farcs2 yr$^{-1}$, 0\farcs3 yr$^{-1}$ and 0\farcs15 yr$^{-1}$ for RW Aur A, RY Tau and DG Tau, respectively, from the latest epoch of the observations. We identify at least five peaks for the RW Aur A jet (labeled as A, B, D, E, F in Figure \ref{fig:rw}), and at least eight peaks in the RY Tau jet (A to H in Figure \ref{fig:ry}). The presence of these knots in the  RW Aur A jet has been previously reported by \citet{Takami20} with the same data sets obtained in 2012-2019. With careful analysis, we additionally identify the probable faint knot C in this jet.
The jet knots G and H for the RY Tau jet were reported by \citet{Uyama22} (labeled as A and B in  their paper). For the DG Tau jet, we identify at least two peaks (C-E), another possible component (B) and a large elongated structure downstream, (A) in Figure \ref{fig:dg}.

\begin{figure}[ht!]
\plotone{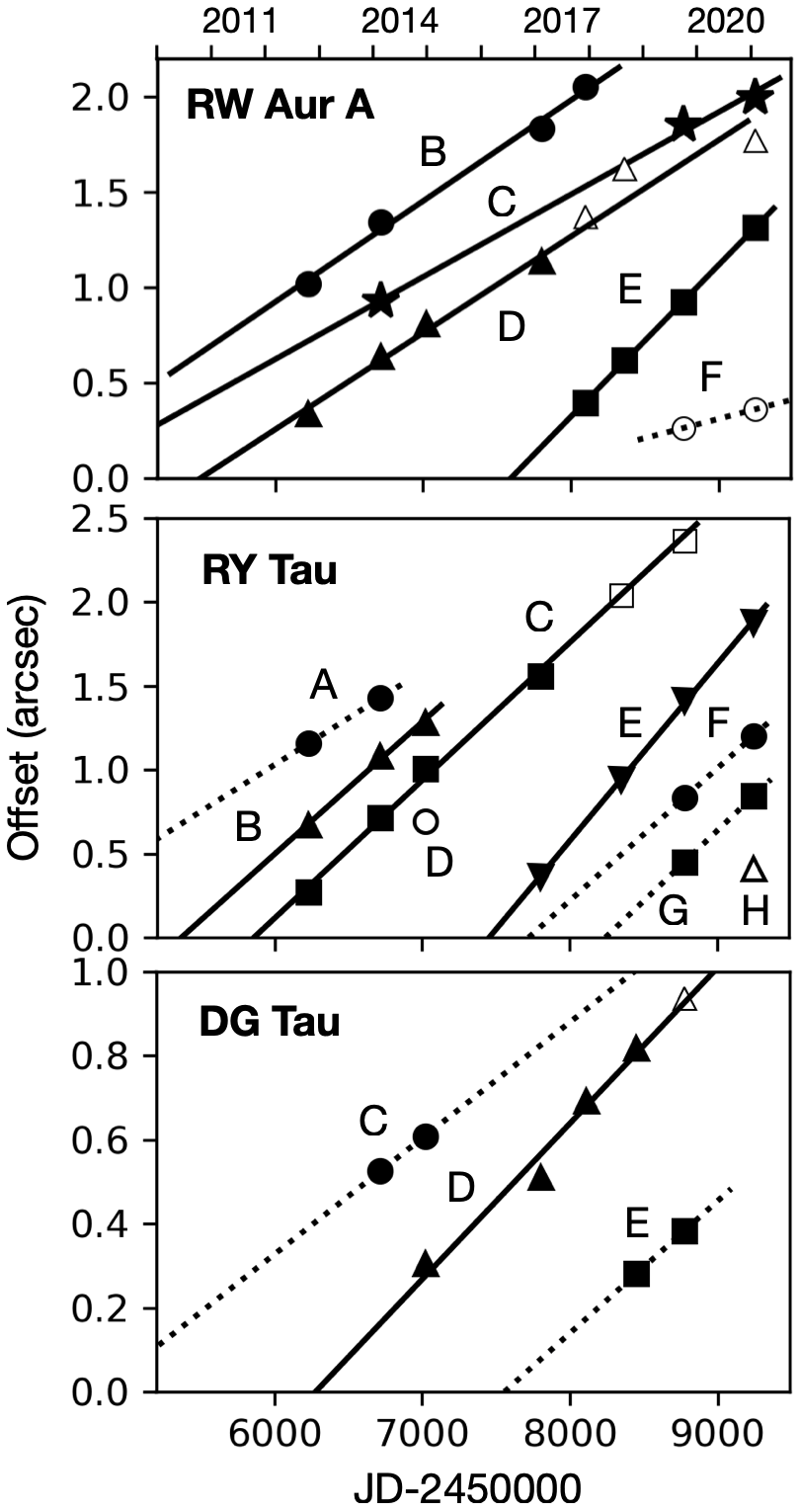}
\caption{Proper motions measured for the individual knots. The dots with filled marks are used for fitting. The solid lines show reliable fitting, while those with dotted lines are tentative due to limited epochs of the observations. The error bar for each dot is not shown as there is no straightforward definition with the given method for the measurements (see text).
\label{fig:pm}}
\end{figure}

Tables \ref{tbl:pm:rw}-\ref{tbl:pm:dg} show the positions of these knots at different epochs measured as follows. We first integrated the intensity distribution across the jet axis and over the velocities for the ranges shown in Table \ref{tbl:log} (i.e., the same ranges as used for the velocity-integrated maps and PV diagrams in Figures \ref{fig:rw}-\ref{fig:dg}). We then applied a polynomial fit for 4-6 positions near the peak and measured the position at the intensity peak. We then fit, for each knot, these offsets from the star as a function of date using a straight line, and derived their proper motions and the date at the origin (Figure \ref{fig:pm}).

There is no straightforward definition for the uncertainty for the measurement of each jet knot position using the above method. We therefore regard the standard deviation of the individual measurements from the fitted straight line as a typical uncertainty, which is also tabulated for each knot in Tables \ref{tbl:pm:rw}-\ref{tbl:pm:dg}.

To derive the uncertainties, we need at least three epochs of observations. We use the measurements of the earliest four epochs of the observations when available. We have more epochs of the observations for Knot D in the RW Aur A jet and knot C, however, we exclude those in the downstream for better accuracies of measurements for the dates at the origin.

%%%%%%%%%%%%%%%%%%%%%%%%
%%% Table : Knot positions and proper motions
%\begin{rotatetable}

\begin{table*}
\caption{Measured Positions and Proper Motions for Jet Knots (RW Aur A) \label{tbl:pm:rw}}
\hspace{-1.8cm}
\begin{tabular}{llcccccc}
\tableline\tableline
&& A & B & C& D & E & F\\ \tableline
Date~~~~~~~	& 2012 Oct 20 	& 1\farcs349	& 1\farcs021	& ---			& 0\farcs339	& ---		& ---	\\
	& ($v_\mathrm{rad}$; km s$^{-1}$)
	& (110$\pm$10) & (86$\pm$10) & --- & (85$\pm$10)  & --- & --- 
	\vspace{0.2cm} \\

	& 2014 Feb 28	& ---		& 1\farcs344	& 0\farcs928	& 0\farcs638	& ---			& ---	\\
	& ($v_\mathrm{rad}$; km s$^{-1}$) 
	& --- & (79$\pm$10) & (72$\pm$10) & (79$\pm$10)  & --- & --- 
	\vspace{0.2cm} \\

	& 2014 Dec 29	& ---		& ---			& ---			& 0\farcs808	& ---			& ---	\\
	& ($v_\mathrm{rad}$; km s$^{-1}$) 
	& --- & --- & --- & (77$\pm$10)  & --- & --- 
	\vspace{0.2cm} \\

	& 2017 Feb 15	& ---		& 1\farcs833	& ---			& 1\farcs136	& ---			& --- \\ 
	& ($v_\mathrm{rad}$; km s$^{-1}$) 
	& --- & (78$\pm$10) & --- & (86$\pm$10)  & --- & --- 
	\vspace{0.2cm} \\

	& 2017 Dec	& ---		& 2\farcs053	& ---			& 1\farcs372	& 0\farcs394	& --- \\
	& ($v_\mathrm{rad}$; km s$^{-1}$) 
	& --- & ---\tablenotemark{a} & --- & ---\tablenotemark{a}  & (128$\pm$20) & --- 
	\vspace{0.2cm} \\

	& 2018 Aug-Sep & ---	& ---			& ---			& 1\farcs626	& 0\farcs620	& --- \\ 
	& ($v_\mathrm{rad}$; km s$^{-1}$) 
	& --- & --- & --- & (78$\pm$10)  & (122$\pm$10) & --- 
	\vspace{0.2cm} \\

	& 2019 Oct 07	& ---		& ---			& 1\farcs851	& ---			& 0\farcs925	& 0\farcs263 \\
	& ($v_\mathrm{rad}$; km s$^{-1}$) 
	& --- & --- & (78$\pm$10) & --- & (114$\pm$10) & (112$\pm$10)
	\vspace{0.2cm} \\

	& 2021 Feb 03	& ---		& ---		&	 1\farcs996	& 1\farcs771	& 1\farcs314	& (0\farcs361)\tablenotemark{b} \\
	& ($v_\mathrm{rad}$; km s$^{-1}$) 
	& --- & --- & ---\tablenotemark{a} & ---\tablenotemark{a} & (87$\pm$20) & (98$\pm$20)\tablenotemark{b}
	\vspace{0.2cm} \\

\tableline
%\multicolumn{2}{l}{Proper Motion in arcsec yr$^{-1}$}	
			$v_{\mathrm tan}$
			& (arcsec yr$^{-1}$)
			& --- 					% A
			& 0.193$\pm$0.011 		% B
			& 0.158$\pm$0.009		% C
			& 0.215$\pm$0.004 		% D
			& 0.290$\pm$0.004 		% E
			& (0.08)\tablenotemark{b} \\	% F
%\multicolumn{2}{l}{~~~~~~~~~~~~~~~~~~~~~in km s$^{-1}$}	
			& (km s$^{-1}$)
			& --- 					% A
			& 143$\pm$9 			% B
			& 117$\pm$7			% C
			& 159$\pm$4			% D
			& 215$\pm$4 			% E
			& ($\sim$60)\tablenotemark{b} \\	% F			
\multicolumn{2}{l}{JD-2450000 at Origin}
			& --- 					% A
			& 4251$\pm$175 		% B
			& 4555$\pm$214		% C
			& 5640$\pm$21 		% D
			& 7595$\pm$14 		% E
			& --- \\				% F
\multicolumn{2}{l}{Interval from Last Ejection (days)}
			& --- 					% A
			& --- 					% B
			& 300$\pm$300		% C
			& 1100$\pm$200 		% D
			& 1955$\pm$25 		% E
			& --- \\				% F
\multicolumn{2}{l}{Fitting Error (arcsec)}
			& --- 					% A
			& 0.038 				% B
			& 0.032				% C
			& 0.005 				% D
			& 0.007 				% E
			& --- \\				% F
\tableline
\end{tabular} \\
\tablenotetext{a}{Signal-to-noises are too low for reliable measurements.}
\tablenotetext{b}{Not currently reliable as it is not clear if the tabulated position for 2021 Feburuary 03 is for knot F observed in 2019 October 7. See text for details.}
\end{table*}

%%% RY

\begin{table*}
\caption{Measured Positions and Proper Motions for Jet Knots (RY Tau) \label{tbl:pm:ry}}
\hspace{-2.5cm}
\begin{tabular}{llcccccccc}
\tableline\tableline
&& A & B & C& D & E & F & G & H\\ \tableline
Date~~~~~~~	& 2012 Oct 27 	& 1\farcs157	& 0\farcs670	& 0\farcs272	& ---		& ---		& ---	 	& ---		& ---	\\
	& ($v_\mathrm{rad}$; km s$^{-1}$) 
	& (--92$\pm$10) & (--99$\pm$10) & (--120$\pm$10) & --- &--- & --- &--- & ---
	\vspace{0.2cm} \\

	& 2014 Feb 28	& 1\farcs428	& 1\farcs084	& 0\farcs716	& ---		& ---		& ---	 	& ---		& ---	\\
	& ($v_\mathrm{rad}$; km s$^{-1}$) 
	& (--104$\pm$10) & (--100$\pm$10) & (--122$\pm$10) & --- &--- & --- &--- & ---
	\vspace{0.2cm} \\

	& 2014 Dec 29	& ---			& 1\farcs285	& 1\farcs006	& 0\farcs690	& ---		& ---		& ---		& ---	\\
	& ($v_\mathrm{rad}$; km s$^{-1}$) 
	& --- & (--95$\pm$10) & (--113$\pm$10) & (--117$\pm$10) &--- & --- &--- & ---
	\vspace{0.2cm} \\

	& 2017 Feb 18	& ---			& ---			& 1\farcs560	& ---		& 0\farcs360	& --- 		& ---		& ---	\\ 
	& ($v_\mathrm{rad}$; km s$^{-1}$) 
	& --- & --- & (--106$\pm$10) & --- & (--93$\pm$10) & --- &--- & ---
	\vspace{0.2cm} \\

	& 2018 Aug 17 & ---			& ---			& 2\farcs038	& ---		& 0\farcs942	& ---		& ---		& --- \\ 
	& ($v_\mathrm{rad}$; km s$^{-1}$) 
	& --- & --- & (--103$\pm$10) & --- & (--97$\pm$10) & --- &--- & ---
	\vspace{0.2cm} \\

	& 2019 Oct 23	& ---			& ---			& 2\farcs365	& ---		& 1\farcs414	& 0\farcs833 	& 0\farcs450	& --- \\
	& ($v_\mathrm{rad}$; km s$^{-1}$) 
	& --- & --- & (--114$\pm$10) & --- & (--108$\pm$10) & (--83$\pm$10) &(--81$\pm$10) & ---
	\vspace{0.2cm} \\

	& 2021 Feb 03	& ---			& ---			& ---			& ---		& 1\farcs878	& 1\farcs202 	& 0\farcs843	& 0\farcs413\\
	& ($v_\mathrm{rad}$; km s$^{-1}$) 
	& --- & --- & --- & --- & (--102$\pm$20) & (--82$\pm$20) & (--81$\pm$20) & (--83$\pm$20)
	\vspace{0.2cm} \\

\tableline
%\multicolumn{2}{l}{Proper Motion in arcsec yr$^{-1}$}	
		$v_{\mathrm tan}$
		& (arcsec yr$^{-1}$)
		& 0.203\tablenotemark{a} 	% A
		& 0.287$\pm$0.018		% B
		& 0.300$\pm$0.016		% C
		& --- 					% D
		& 0.385$\pm$0.005		% E
		& 0.288\tablenotemark{a}	% F
		& 0.307\tablenotemark{a}	% G
		& ---					% H
		\\
%\multicolumn{2}{l}{~~~~~~~~~~~~~~~~~~~~~in km s$^{-1}$}	
		& (km s$^{-1}$)
		& 120\tablenotemark{a} 	% A
		& 170$\pm$13		% B
		& 178$\pm$12		% C
		& --- 					% D
		& 228$\pm$7		% E
		& 171\tablenotemark{a}	% F
		& 182\tablenotemark{a}	% G
		& ---					% H
		\\
\multicolumn{2}{l}{JD-2450000 at Origin}	
		& 4145\tablenotemark{a} 	% A
		& 5367$\pm$85 		% B
		& 5859$\pm$66 		% C
		& ---					% D
		&7455$\pm$17			% E
		&7724\tablenotemark{a} 	% F
		&8244\tablenotemark{a} 	% G
		& --- 					% H
		\\
\multicolumn{2}{l}{Interval from Last Ejection (days)}
		& --- 					% A
		& $\sim$1200 			% B
		& 500$\pm$100 		% C
		& ---					% D
		&(1600$\pm$70)\tablenotemark{b}		% E
		&$\sim$300 			% F
		&$\sim$500 			% G
		& --- 					% H
		\\
\multicolumn{2}{l}{Fitting Error (arcsec)}
		& --- 					% A
		& 0.020 				% B
		& 0.041				% C
		& --- 					% D
		& 0.013 				% E
		& ---					% F
		& ---					% G
		& --- \\				% H
\tableline
\end{tabular} \\
\tablenotetext{a}{Uncertainties are not clear due to limited epochs of the observations.}
\tablenotetext{b}{Interval from the second last ejection (C), as we were not able to measure that for the last ejection (D).}
\end{table*}

%%% DG

\begin{table}
\caption{Measured Positions and Proper Motions for Jet Knots (DG Tau) \label{tbl:pm:dg}}
\hspace{-2cm}
\begin{tabular}{llccc}
\tableline\tableline
&& C& D & E \\ \tableline
Date~~~~~~~	& 2013 Feb 09 	& ---	& ---	& --- \\
	& ($v_\mathrm{rad}$; km s$^{-1}$) 
	& --- & --- & --- % & (--85$\pm$10) 
	\vspace{0.2cm} \\

	& 2014 Feb 28	& 0\farcs525	& ---			& --- \\
	& ($v_\mathrm{rad}$; km s$^{-1}$) 
	& (--142$\pm$10) & --- & --- % & (--85$\pm$10) 
	\vspace{0.2cm} \\

	& 2014 Dec 29	& 0\farcs609	& 0\farcs305	& --- \\
	& ($v_\mathrm{rad}$; km s$^{-1}$) 
	& (--134$\pm$10) & (--120$\pm$10) & --- % & (--85$\pm$10) 
	\vspace{0.2cm} \\

	& 2017 Feb 17	& ---			& 0\farcs510	& --- \\ 
	& ($v_\mathrm{rad}$; km s$^{-1}$) 
	& --- & (--152$\pm$10) & --- % & (--85$\pm$10) 
	\vspace{0.2cm} \\

	& 2017 Dec	& ---			& 0\farcs693	& --- \\
	& ($v_\mathrm{rad}$; km s$^{-1}$) 
	& --- & (--134$\pm$20) & --- % & (--85$\pm$10) 
	\vspace{0.2cm} \\

	& 2018 Nov 27 & ---			& 0\farcs818	& 0\farcs281 \\ 
	& ($v_\mathrm{rad}$; km s$^{-1}$) 
	& --- & (--136$\pm$10) & (--141$\pm$10) % & (--85$\pm$10) 
	\vspace{0.2cm} \\

	& 2019 Oct 17	& ---			& 0\farcs938	& 0\farcs383 \\
	& ($v_\mathrm{rad}$; km s$^{-1}$) 
	& --- & (--140$\pm$10) & (--155$\pm$10) % & (--85$\pm$10) 
	\vspace{0.2cm} \\

\tableline
%\multicolumn{2}{l}{Proper Motion in arcsec yr$^{-1}$}	
		$v_{\mathrm tan}$
		& (arcsec yr$^{-1}$)
		& 0.101\tablenotemark{a} 			% C
		& 0.136$\pm$0.016 	% D
		& 0.115\tablenotemark{a} \\			% E
%\multicolumn{2}{l}{~~~~~~~~~~~~~~~~~~~~~in km s$^{-1}$}	
		& (km s$^{-1}$)
		& 66\tablenotemark{a} 			% C
		& 89$\pm$13 	% D
		& 75\tablenotemark{a} \\			% E
\multicolumn{2}{l}{JD-2450000 at Origin}	
		& 4816\tablenotemark{a}			% C
		& 6279$\pm$197 	% D
		& 7557\tablenotemark{a} \\			% E
\multicolumn{2}{l}{Interval from Last Ejection (days)}
		& ---			% C
		& $\sim$1500 	% D
		& $\sim$1300 \\	 % E
\multicolumn{2}{l}{Fitting Error (arcsec)}	
		& --- 				% C
		& 0.038 			% D
		& --- 	\\			% E
\tableline
\end{tabular} \\
\tablenotetext{a}{Uncertainties are not clear due to limited epochs of the observations.}
\end{table}

%%%%%%%%%%%%%%%%%%%%%%%%
%%% Figure: Proper motions
%%%%%%%%%%%%%%%%%%%%%%%%

One might suspect that the measured knot positions are affected by the selected spatial range across the jet axis and the velocity range. To investigate this, we have derived knot positions by increasing/decreasing each range by 30 \%. These analysis yielded a typical positional difference of 0\farcs007, which has little effect on the fitting parameters tabulated at the bottom of Tables \ref{tbl:pm:rw}-\ref{tbl:pm:dg}. The changes in the fitting parameters with this analysis are significantly smaller than the uncertainties shown in Tables \ref{tbl:pm:rw}-\ref{tbl:pm:dg}.

As shown in Tables \ref{tbl:pm:rw}-\ref{tbl:pm:dg} and Figure \ref{fig:pm}, the knots in each jet would have different tangential velocities for the RW Aur A and RY Tau jets: these are, 0\farcs16-0\farcs29  yr$^{-1}$ (knots B-E; corresponding to 120-220 km s$^{-1}$) and 0\farcs2-0\farcs4 yr$^{-1}$ (knots ABCEFG; corresponding to 120-240 km s$^{-1}$), respectively. For the DG Tau jet, the epochs of the observations are not sufficient for investigating the variations of the tangential velocities between the knots.

For knot F in the RW Aur jet, one may infer a proper motion of 0\farcs08 yr$^{-1}$ based on their peak positions. This value is significantly smaller than the others (0\farcs19-0\farcs29 yr$^{-1}$) (See Figure \ref{fig:pm} and Table \ref{tbl:pm:rw}). As shown in Figure \ref{fig:rw}, this knot looks elongated in the latest epoch (2021 February 3). We cannot exclude the possibility that another new knot emerged and it is apparently near the original knot F in this epoch, making a single elongated knot-like structure in the images with the given angular resolution. Observations for another later epoch would allow us to prove or reject this explanation.

The knots with large proper motions may have collided, or may collide in the future, with others downstream.
Figure \ref{fig:ry} show that Knot D from RY Tau may have collided with Knot C in 2016-2017, but it is not conclusive due to insufficient epochs of measurements for D.
Figure \ref{fig:pm} suggests that Knots CDE from RW Aur A will collide in the next several years at $\sim$3\arcsec~from the star.

In Tables \ref{tbl:pm:rw}-\ref{tbl:pm:dg} we also list the epochs of jet knot ejections at the star measured based on proper motions measurements, and also their time invervals.
The jet knot ejections from RW Aur A and RY Tau show irregular time intervals between 300--2000 and 300--1200 days, respectively. We do not have sufficient epochs of observations to investigate whether the time intervals are irregular or not for the DG Tau jet. However, the measured intervals of 1300-1500 days are different from those inferred from the previous studies: $\sim$1800 or $\sim$900 days, or a combination of these two, between 1980 and 2005 \citep{Pyo03,Agra11,Rodriguez12,White14a}. This discrepancy suggests that the jet knot ejections are irregular over a timescale of $\sim$40 years.

\begin{figure*}[ht!]
%\plotone{I_x.png}
\includegraphics[width=18cm]{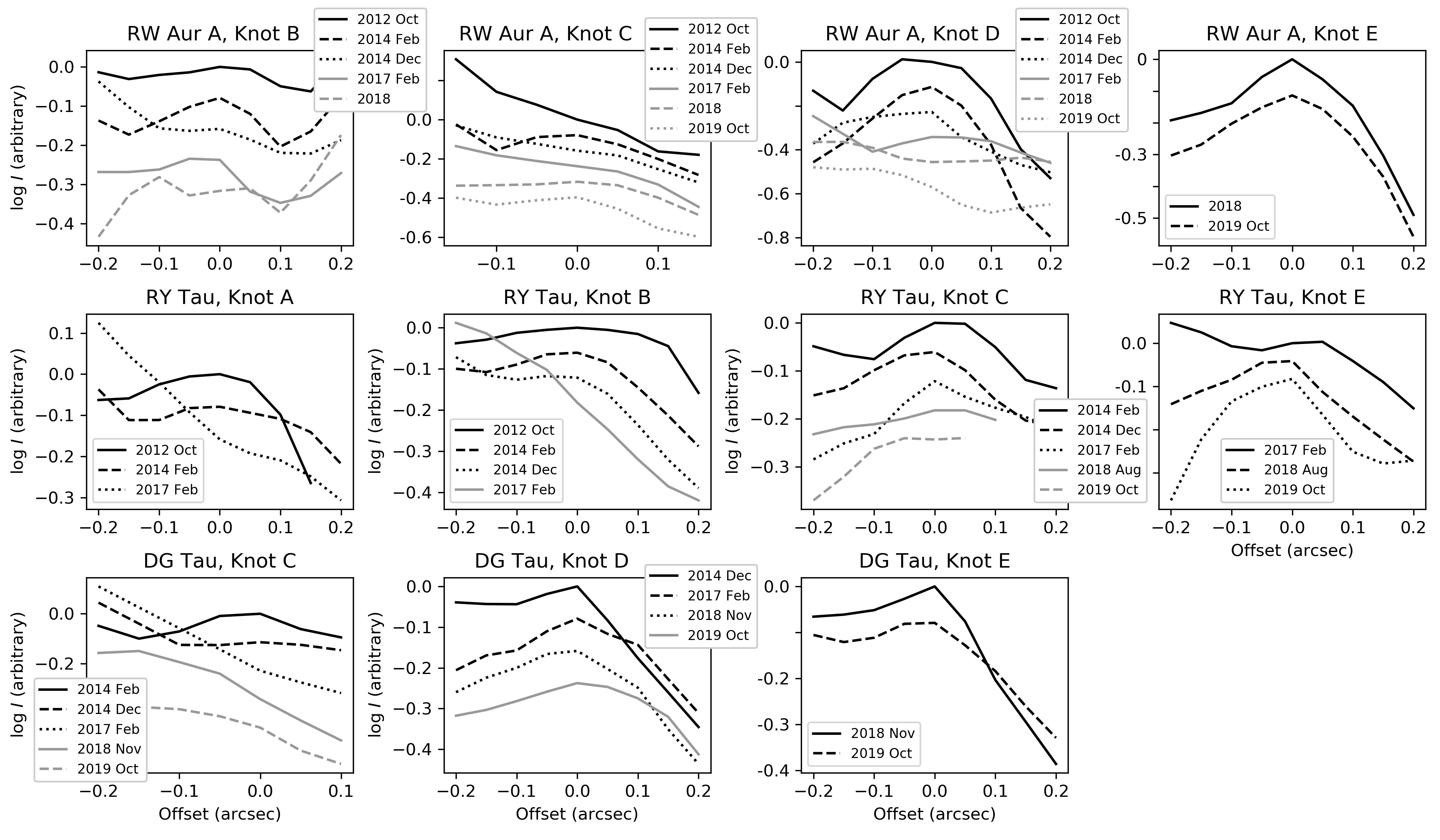}
\caption{One-dimensional intensity distributions along the jet axis at the individual knots. We arbitrarily scale the intensity distributions and show them in chronological order from top to bottom for each panel organized for the individual knots. Each of the observed distributions is shown using a black/gray solid/dashed/dotted curve in terms of the measured peak position. For some distributions without a peak (see also Tables \ref{tbl:pm:rw}-\ref{tbl:pm:dg}), we determined their zero-positions in the plot using the fitted lines of the knot positions with the other epochs of the observations (Figure \ref{fig:pm}).
\label{fig:I_x}}
\end{figure*}

Figure \ref{fig:I_x} shows how the observed intensity distribution along the jet axis changes with time at the positions of the individual knots. The data obtained using SINFONI and OSIRIS are not included as higher angular resolutions (see Table \ref{tbl:log}) makes it more difficult to investigate the actual time variations of the spatial structures.  As for the measurements of the peak positions, we spatially integrated the intensity across the jet axis. We then arbitrarily scale the intensity distributions and show them in chronological order from top to bottom for each panel organized for each knot. For some knots and epochs, we were not able to measure the peak positions using the method described above. For those, we adopt the values based on the proper motion measurements.

%%%%%%%%%%%%%%%%%%%%%%%%
%%% Figure: I_x for knots
%%%%%%%%%%%%%%%%%%%%%%%%

Figure \ref{fig:I_x} shows that the knots have a spatial extent along the jet comparable to the angular resolution (typically 0\farcs15 for the plotted data; see Table \ref{tbl:log}). Many of these knots are smeared for later epochs: these are, knot E from RW Aur A; knot ABC from RY Tau; and knot CD from DG Tau. Such a trend is less clear for the remaining knots.

%%%%%%%%%%%%%%%%%%%%%%%%%%%%%%%%%%%%%%%%
%%%%%%%%%%%%%%%%%%%%%%%%%%%%%%%%%%%%%%%%
%%% 3.2, spatial distributions along the jet axis
%%%%%%%%%%%%%%%%%%%%%%%%%%%%%%%%%%%%%%%%
%%%%%%%%%%%%%%%%%%%%%%%%%%%%%%%%%%%%%%%%

\subsection{Spatial Structures across the Jet Axis} \label{results:y}

For all jets, the emission across the jet axis is marginally resolved, with full width half maxima (FWHMs) up to 0\farcs4--0\farcs5, 2-3 times larger than the angular resolution. These FWHM values are similar to previous observations of the same jets with the same emission line \citep{Agra11,Garufi19} and optical emission lines \citep{Dougados00,Bacciotti00,Woitas02,Agra09,Liu12,Garufi19}.
The [\ion{Fe}{2}] 1.644-\micron~emission in our velocity-integrated maps shows a gaussian-like or a symmetric triangular distribution except for the structure A in the DG Tau jet, for which we find asymmetric profiles in some positions. In Figure \ref{fig:dg}, the spatial distribution of this structure is similar to an asymmetric bow shock modeled by \citet{Raga01}.

Some line profiles are associated with faint and more extended emission, in particular associated with bright jet regions. While this could be real emission at the observed positions, we cannot currently exclude the possibility that these are due to halos associated with the PSFs.

For RW Aur A and DG Tau, the observed knots are closely spatially aligned along a single PA. In contrast, Figure \ref{fig:ry} shows recognizable offsets from a single jet axis for the RY Tau jet. The most remarkable offsets are seen for knot E observed in 2019 and 2021,  about $\sim$0\farcs1 offset from the jet axis shown in Figure \ref{fig:ry}. The PA of knot E from the star is different from knot G by $\sim$8\arcdeg. In 2014, we also identify a small offset for B, whose PA is different from that of C by 3\arcdeg-6\arcdeg. For RW Aur, we also identify a marginal offset for knot B observed in 2012 and early 2014. In the latter epoch, the PA of the knot B is different from A by $\sim$2\arcdeg.

These differences in directions observed in the RY Tau jet may be due to wiggling motions of the jet \citep{Lavalley00,Raga01,Garufi19,Uyama22}.
In particular, \citet{Garufi19} observed the RY Tau jet over a $\sim$6\arcsec~scale in 2017 and 2019, and identified a similar pattern. These authors measured the jet PA of 290\arcdeg~at the base of the jet and an elongated structure at 5\arcsec-6\arcsec~away from the star, and 295\arcdeg~for a knot 3\arcsec-4\arcsec~away from the star. This pattern is explained as an outer extension of the jet wiggling pattern seen in our 2017 and 2019 images, in which the jet shows a smaller PA between knots C and F (Figure \ref{fig:ry}).

\citet{Garufi19} discussed the following two possibilities for the origin of the jet wiggling using the observed spatial distributions of the jet and its velocity: (1) orbital motion of the primary star induced by a stellar companion \citep[e.g.,][]{Anglada07}; and (2) precession of the inner disk (i.e., where the jet is launched) induced by a substellar companion, whose orbit is misaligned with the outer disk plane \citep[e.g.,][]{Zhu19}. \citet{Garufi19} excluded the first scenario because the companion required to explain the observed jet wiggling ($M_*$=1.1 $M_\sun$, $d$$\sim$0\farcs1) has not been detected by high-resolution imaging observations to date (Section \ref{targets:ry}). These authors demonstrated that the second scenario would work: a substellar companion to explain the observed jet wiggling would be too faint and/or too close to the primary star to be detected by these observations. The alternative scenario, which was not discussed by \citet{Garufi19}, is that the precession of the disk is induced by magnetic torques associated with the jet/outflow, and a resultant warping instability in the inner disk \citep{Lai03,Erkal21a}.

%%%%%%%%%%%%%%%%%%%%%%%%%%%%%%%%%%%%%%%%
%%%%%%%%%%%%%%%%%%%%%%%%%%%%%%%%%%%%%%%%
%%% 4.4 Peak intensities and intensity ratios
%%%%%%%%%%%%%%%%%%%%%%%%%%%%%%%%%%%%%%%%
%%%%%%%%%%%%%%%%%%%%%%%%%%%%%%%%%%%%%%%%

\subsection{Peak Intensities and Intensity Ratios} \label{results:I}

\begin{figure*}[ht!]
%\plotone{photometry.png}
\includegraphics[width=18cm]{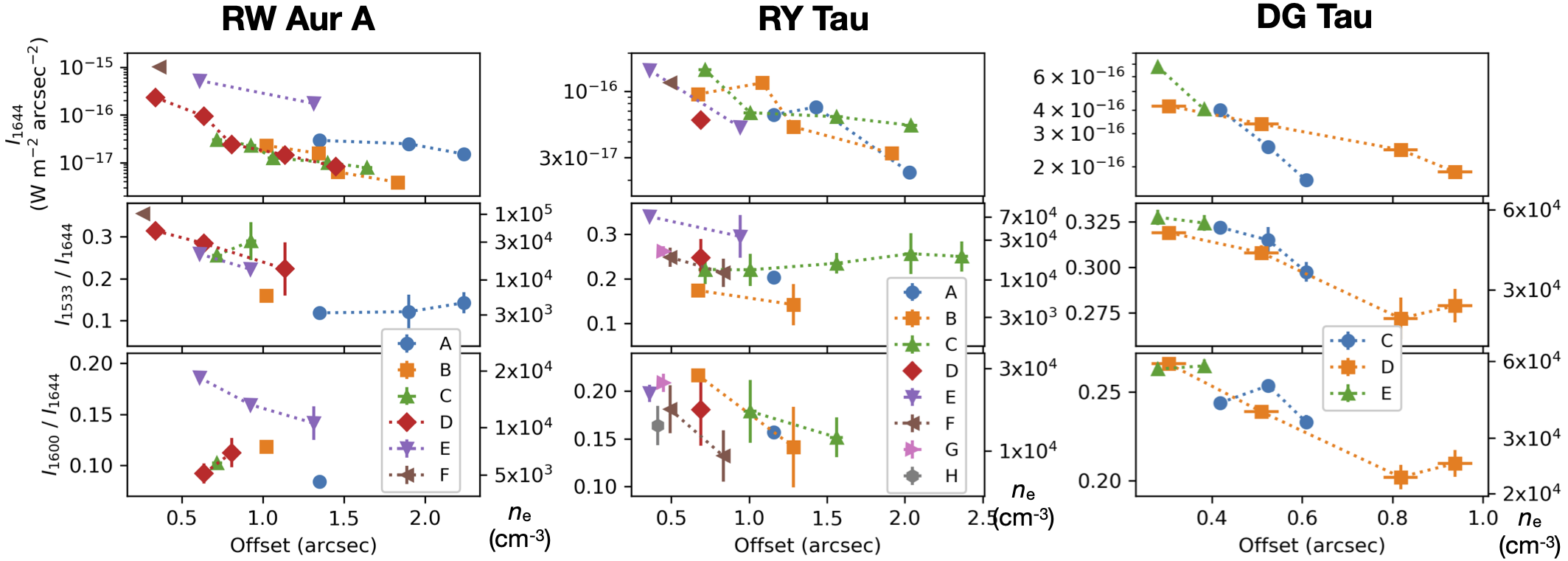}
\caption{The [\ion{Fe}{2}] 1.644 \micron~peak intensities (top), the $I_{1.533\micron}/I_{1.644\micron}$ intensity ratios (middle), and  the $I_{1.600\micron}/I_{1.644\micron}$ intensity ratios as a function of the distance to the star. See text for the selection criteria of the plotted data. 
For the intensity ratios, we show the corresponding electron densities for $T_\mathrm{e} = 10^4$ K at the right side of the individual panels. Some dots are associated with the horizontal error bars based on uncertainties tabulated in Tables \ref{tbl:pm:rw}-\ref{tbl:pm:dg}. The vertical error bars are shown only for those larger than the dots. See text for other details.
\label{fig:Ipeak_vs_x}}
\end{figure*}

\begin{figure*}[ht!]
%\plotone{photometry.png}
\includegraphics[width=18cm]{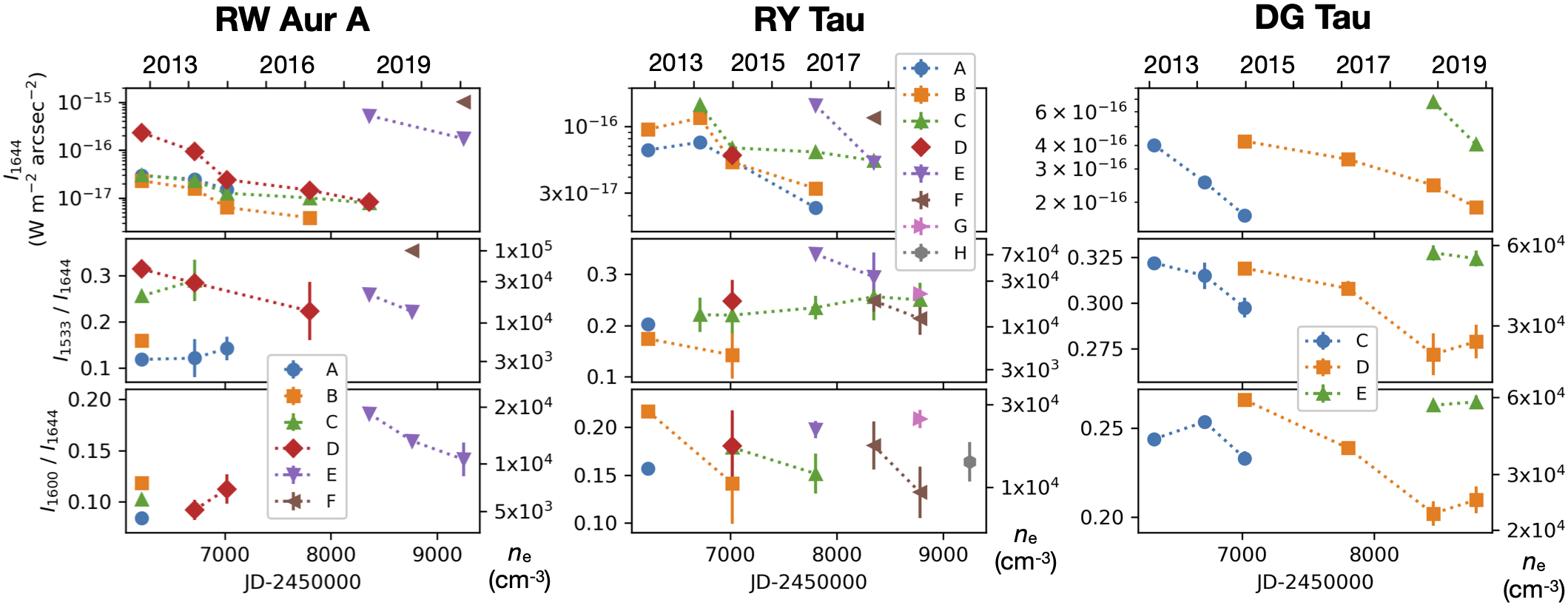}
\caption{Same as Figure \ref{fig:Ipeak_vs_x} but with JD for the horizontal axis. In the top panels we show the corresponding calendar years.
\label{fig:Ipeak_vs_JD}}
\end{figure*}

%For this subsection use data obtained with the photometric conditions only (see Table \ref{tbl:log}).
Figures \ref{fig:Ipeak_vs_x} and \ref{fig:Ipeak_vs_JD} show the 1.644-\micron~peak intensities of the individual knots as a function of the projected distance to the star and the observed date, respectively. To minimize the effect of different angular resolutions, we measured the peak intensity for each knot in a 0\farcs15$\times$0\farcs15 area. In the same figures we also plot the 1.533/1.644-\micron~and 1.600/1.644-\micron~intensity ratios at the 1.644-\micron~intensity peaks. For these figures we selected the data points for each line with the following criteria: (1) we identify the intensity peaks in the velocity-integrated maps; and (2) the measured intensities or intensity ratios are above 3-$\sigma$ levels. Furthermore, we use the data obtained with photometric conditions only for the 1.644-\micron~peak intensities.

%%%%%%%%%%%%%%%%%%%%%%%%
%%% Figure: I_peak_vs_x

%%%%%%%%%%%%%%%%%%%%%%%%
%%% Figure: I_peak_vs_JD
%%%%%%%%%%%%%%%%%%%%%%%%

In Figures \ref{fig:Ipeak_vs_x} and \ref{fig:Ipeak_vs_JD}, the peak intensities of the individual knots tend to decrease as the Julian date and the distance to the star increase, by a factor of up to $\sim$30 during the time of our observations. In Figure \ref{fig:Ipeak_vs_x}, the peak intensities between different knots show a fairly good correlation for many knots from RW Aur A, and all of those from RY Tau. In the top panels of Figure \ref{fig:Ipeak_vs_JD}, the peak intensities measured for the individual knots show different inclinations, i.e., different timescales for intensity decays, in particular for the jet knots from RY Tau and DG Tau. In the case of Knot D from RW Aur A, the peak intensity decreases by a factor of $\sim$10 during the first $\sim$800 days, but by a factor of $\sim$3 during the subsequent $\sim$1300 days. In contrast, the measured peak intensities in the RY Tau jet marginally increased for knot A and knot B during JD=2456228 to 2456716, both by a factor of $\sim$1.2.

\begin{figure}[ht!]
%\plotone{photometry.png}
\includegraphics[width=7cm]{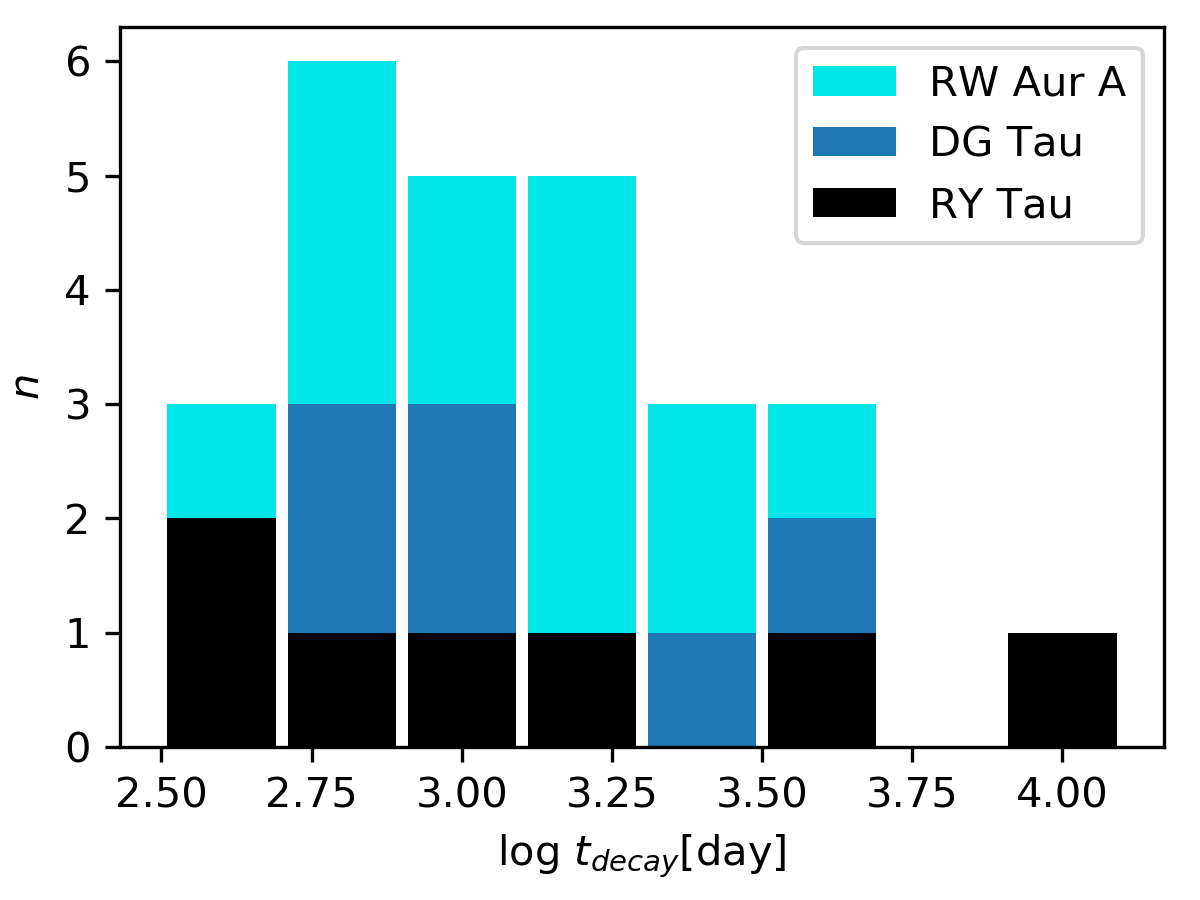}
\caption{The number distributions of the decay timescales for the [\ion{Fe}{2}] 1.644 \micron~intensity peaks. See text for details.
\label{fig:t_decay}}
\end{figure}

To further discuss the decay timescale of the emission, we define the timescale $t_{\mathrm{decay}}$ based on the following equation:
%%%
%
\begin{equation}
I_1 = I_0~\mathrm{exp}(-t/t_{\mathrm{decay}}),
\label{eq:t_decay}
\end{equation}
%
%%%
where $I_0$ and $I_1$ are the peak intensities of a knot for two subsequent epochs of observation; and $t$ is the time interval of these epochs.
Figure \ref{fig:t_decay} shows their number distributions, except the cases for which the peak intensities marginally increased. Most of these are distributed between 300 and 3600 days, with a median value of $\sim$1000 days.

%%%%%%%%%%%%%%%%%%%%%%%%
%%% Figure: I_peak_vs_JD
%%%%%%%%%%%%%%%%%%%%%%%%

In Figures \ref{fig:Ipeak_vs_x} and \ref{fig:Ipeak_vs_JD}, we find observed 1.533/1.644-\micron~and 1.600/1.644-\micron~intensity ratios of 0.1--0.35 and  0.08--0.27, respectively. At the right side of each panel for the intensity ratio, we mark the corresponding electron densities calculated by \citet{Nisini02,Pesenti04,Takami06a}. These indicate that the electron densities at the intensity peaks range between $3 \times 10^3$ and $1 \times 10^5$ cm$^{-3}$. The line ratios and electron densities measured in the DG Tau and RW Aur A jets are similar to the previous observations for the same jets \citep{Lavalley00,Bacciotti00,Woitas02,Dougados02,Coffey08,Melnikov09,Agra11,White14a}.

In Figure \ref{fig:Ipeak_vs_x}, the intensity ratios (and therefore the electron densities) decrease downstream for the DG Tau jet, however, this trend is not very clear for the others, perhaps because of modest signal-to-noise. Again, this trend was also measured before for the DG Tau jet by  \citet{Lavalley00,Dougados02,White14a}. In Figure \ref{fig:Ipeak_vs_JD}, the peak intensities of knot D from DG Tau appear to decrease with time. While a similar trend is observed for some other knots from all the stars, better signal-to-noise is required for confirmation. 

\begin{figure*}[ht!]
%\plotone{photometry.png}
\includegraphics[width=18cm]{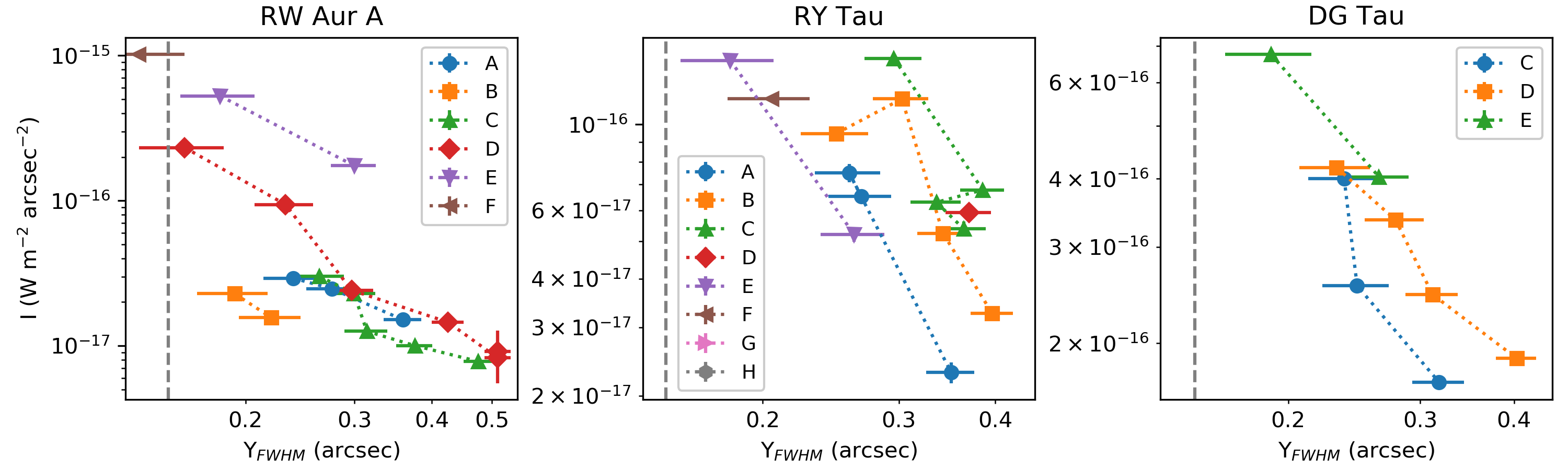}
\caption{Correlation between the FWHM across the jet and the peak intensities for the knots in the [\ion{Fe}{2}] 1.644 \micron~emission. The FWHMs are measured without deconvolution of the observed PSF. The vertical error bars are shown only for those larger than the dots. The vertical dashed line at the left of each panel shows a typical angular resolution (0\farcs15).
\label{fig:FWHM_vs_I}}
\end{figure*}

Figure \ref{fig:FWHM_vs_I} shows a correlation between the peak intensity and the FWHM of the jet width for the individual knots. We use data obtained in photometric conditions only, as for the plots for the peak intensities in Figures \ref{fig:Ipeak_vs_x} and \ref{fig:Ipeak_vs_JD}. We have not applied the deconvolution procedure, and we indicate a typical PSF size (0\farcs15; see Table \ref{tbl:log}) using the vertical black dashed lines in the individual panels. In the figure the peak intensity and the FWHM show a negative correlation, implying that the compact knots show larger surface brightnesses. For the jets from RW Aur A and RY Tau, the measured ranges for the FWHMs (a factor of 2--3) are significantly smaller than those for the peak intensities (a factor of 10--100). In contrast, those ranges are similar for the jet from DG Tau: factors of $\sim$2 and $\sim$3 for the FWHMs and the peak intensities, respectively.

%%%%%%%%%%%%%%%%%%%%%%%%
%%% Figure: I_peak_vs_JD
%%%%%%%%%%%%%%%%%%%%%%%%

%%%%%%%%%%%%%%%%%%%%%%%%%%%%%%%%%%%%%%%%
%%%%%%%%%%%%%%%%%%%%%%%%%%%%%%%%%%%%%%%%
%%% 3.3, velocities along the jet axis
%%%%%%%%%%%%%%%%%%%%%%%%%%%%%%%%%%%%%%%%
%%%%%%%%%%%%%%%%%%%%%%%%%%%%%%%%%%%%%%%%

\subsection{Radial vs. Tangential Velocities} \label{results:v}

The radial velocity profiles at the individual positions in the jet can be reasonably well fitted by a single gaussian with FWHMs comparable to the instrument resolution for all the jets and a majority of epochs. In the jets from DG Tau and RY Tau, the FWHMs reach up to 120-140 km s$^{-1}$ close to the star in some epochs. Furthermore, some line profiles are associated with faint wing emission at low velocities probably due to one or more of the following: (1) a slow wide-angled wind \citep[see][for a review]{Eisloffel00_PPIV}; (2) an onion-like kinematic structure in the jet, with a highly collimated central flow surrounded by slower components \citep{Bacciotti00}; (3) ambient gas entrained by the jet \citep{Pyo03,White14a,White16}. We leave detailed analysis of these components as possible future work due to the difficulty of analysis with a limited velocity resolution and their faint nature. Similarly, we leave a search for the spinning motions in the jet \citep[e.g.,][]{Coffey04,Coffey07,Coffey11,Coffey12,Coffey15,Lee17,Erkal21a} as possible future work because this study requires careful analysis of radial velocities significantly smaller than the instrument resolutions.

Using gaussian fitting, we derive velocities in the jet of 70 to 130 km s$^{-1}$ for RW Aur A, --70 to  --120 km s$^{-1}$ for RY Tau and --120 to --200 km s$^{-1}$ for DG Tau. These spatially vary in the jet, suggesting that the jet launching velocities vary on timescales of a few years or longer, as for the tangential velocities we discussed in Section \ref{results:x}. In Tables \ref{tbl:pm:rw}-\ref{tbl:pm:dg} we list the radial velocities measured at the peak positions of the knots.
We do not find any clear evidence for time variation of the radial velocities for any knot. 
For knot E in the RW Aur jet, the radial velocity may have decreased from $\sim$120 to $\sim$90 km s$^{-1}$ between 2017 and 2021, but the difference is still comparable to the uncertainties for the measurements.

\begin{figure*}[ht!]
\plotone{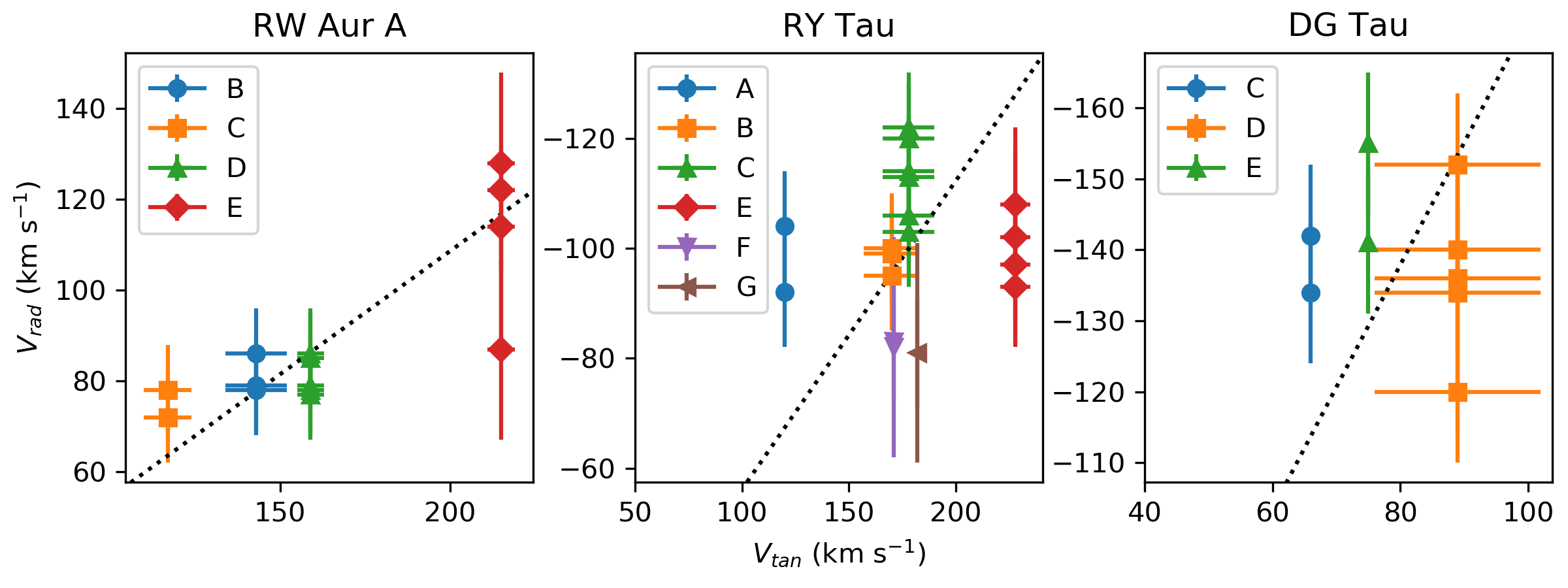}
\caption{Correlations between the tangential and radial velocities for the individual jets. The dashed lines are based on the average of the inclinations of the ejections. The error bars for the tangential velocities are shown only for those we were able to measure (see Section \ref{results:x} and Tables \ref{tbl:pm:rw}--\ref{tbl:pm:dg}).
\label{fig:v_tan_vs_v_rad}}
\end{figure*}

Figure \ref{fig:v_tan_vs_v_rad} shows correlations between the radial velocities and the tangential velocities we measured in Section \ref{results:x}.
%A fairly good correlation between these velocities may have been observed for the RW Aur A jet. 
Large uncertainties in the radial velocities hamper the investigation for whether these two velocities are correlated.
We derive average jet inclination angles of 28\arcdeg$\pm$2\arcdeg, 29\arcdeg$\pm$5\arcdeg, and 59\arcdeg$\pm$3\arcdeg~for the jet from RW Aur A, RY Tau, and DG Tau, respectively.

%%%%%%%%%%%%%%%%%%%%%%%%
%%% Figure: possible stationary shocks
%%%%%%%%%%%%%%%%%%%%%%%%

%%%%%%%%%%%%%%%%%%%%%%%%%%%%%%%%%%%%%%%%
%%%%%%%%%%%%%%%%%%%%%%%%%%%%%%%%%%%%%%%%
%%%%%%%%%%%%%%%%%%%%%%%%%%%%%%%%%%%%%%%%
%%% 5. Discussion
%%%%%%%%%%%%%%%%%%%%%%%%%%%%%%%%%%%%%%%%
%%%%%%%%%%%%%%%%%%%%%%%%%%%%%%%%%%%%%%%%
%%%%%%%%%%%%%%%%%%%%%%%%%%%%%%%%%%%%%%%%

\section{Discussion} \label{discussion}

%Most the results shown in Section \ref{results}, if not all, seem to be consistent with the hypothesis that the jet knots are moving downstream. 
In Section \ref{results}, we presented the results and analysis, tentatively attributing the observed jet knots to `moving knots'.
In Section \ref{discussion:moving} we further discuss the moving knot scenario with their possible heating mechanisms.
In Section \ref{discussion:stationary} we show that some knots in the inner regions may be alternatively attributed to `stationary shocks', rather than moving knots, as discussed in some previous studies.

%%%%%%%%%%%%%%%%%%%%%%%%%%%%%%%%%%%%%%%%
%%%%%%%%%%%%%%%%%%%%%%%%%%%%%%%%%%%%%%%%
%%% 5.1 The Moving-Knot Scenario
%%%%%%%%%%%%%%%%%%%%%%%%%%%%%%%%%%%%%%%%
%%%%%%%%%%%%%%%%%%%%%%%%%%%%%%%%%%%%%%%%

\subsection{The Moving-Knot Scenario} \label{discussion:moving}

In Section \ref{discussion:moving:v} we perform comparisons of the tangential and radial velocities of the jet knots between our analysis in Section \ref{results} and in the literature, and discuss their time variations over up to $\sim$200 years and implications for the moving-knot scenario.
In Section \ref{discussion:moving:shocks} we attempt to attribute the observed trends to shocks, a popular interpretation for jet knots.
In Section \ref{discussion:moving:misc} we discuss alternative explanations for the physical nature of the moving jet knots.
In Section \ref{discussion:moving:acc} we briefly discuss the implications for the regular/irregular time intervals of the jet knot ejections.

%%%%%%%%%%%%%%%%%%%%%%%%%%%%%%%%%%%%%%%%
%%% 5.1.1  The Moving-Knot Scenario

\subsubsection{Long-term Variation of Jet Velocities} \label{discussion:moving:v}

%As shown in Section \ref{results:x}, most of the knots, if not all, can be attributed to moving knots with constant velocities.
In Table \ref{tbl:vt_lit} we compare the measured proper motions between this study and previous work. For the jets from RW Aur A and RY Tau, the proper motions we measured are similar to those in the literature. In contrast, those we measured for the DG Tau jet are smaller than the values in the literature by a factor of $\sim$2.

%%%%%%%%%%%%%%%%%%%%%%%%
%%% Table : Radial Velocities
%\begin{rotatetable}

\begin{table*}
\caption{Proper Motions of Jet Knots\label{tbl:vt_lit}}
\begin{tabular}{lccccccc}
\tableline\tableline
Star  & Proper Motion 	& $V_\mathrm{tan}$	& Angular Distances 	& Lines\tablenotemark{a} & Approximate Years & Reference\tablenotemark{b}\\
	&(arcsec yr$^{-1}$)	& (km s$^{-1}$)		& From the Star	&& of Ejections  \\
	&  				&				&      				&& From the Star		    & 
\\ \tableline
RW Aur A	& 0.15-0.23	& 110-170			&  0\farcs3-20\arcsec	& [\ion{Fe}{2}], [\ion{S}{2}], H$\alpha$ &  1830-2000	& 1\\
		& 0.16-0.24	& 120-180			& 1\arcsec-3\arcsec		& [\ion{S}{2}]					& 1930-2000	& 2 \\
		& {\bf 0.16-0.29} & {\bf 120-210}	& {\bf 0\farcs2-3\arcsec} 	& {\bf [\ion{Fe}{2}]}				& {\bf 2007-2020} & {\bf This work} \\		
RY Tau	& 0.3-0.4		& 180-240			& 1\arcsec-6\arcsec		& [\ion{Fe}{2}]					& 1980-2010	& 3 \\
		& {\bf 0.2-0.4} 	& {\bf 120-240}		& {\bf 0\farcs2-2\farcs5} 	& {\bf [\ion{Fe}{2}]}				& {\bf 2007-2020} & {\bf This work} \\
		& $\sim$0.3	& $\sim$200		& 0\farcs15-1\arcsec		&  [\ion{Fe}{2}], H$\alpha$			& 2019-2021	& 4 \\
DG Tau	& 0.3			& 200			& $\sim$0\farcs3		& [\ion{O}{1}]					& 1985		& 5 \\
		& 0.27-0.3		& 180-200			& 0\farcs1-3\farcs5		& various						& 1995-2000 	& 6 \\
		& 0.17-0.33	& 110-220			& 0\farcs2-1\farcs4		& [\ion{Fe}{2}]					& 1998--2004	& 7 \\
		& {\bf 0.10-0.14} & {\bf 70-90}		& {\bf 0\farcs2-2\arcsec} & {\bf [\ion{Fe}{2}]}				& {\bf 2008-2020} & {\bf This work} \\
		\tableline
\end{tabular}
\tablenotetext{a}{[\ion{Fe}{2}] 1.644 \micron, [\ion{S}{2}] 6731 \AA~and  [\ion{O}{1}] 6300 \AA~for the forbidden lines.}
\tablenotetext{b}{(1) \citet{Berdnikov17};
(2) \citet{Lopez03};
(3) \citet{Garufi19};
(4) \citet{Uyama22};
(5) \citet{Dougados00}
(6) \citet{Pyo03};
(7) \citet{White14a}
}
\end{table*}
%\end{rotatetable}
%%%%%%%%%%%%%%%%%%%%%%%%

In Table \ref{tbl:vr_lit} we compare the radial velocities of the jet line emission with the previous observations at sub-arcsecond resolutions. For this table we include the observations of the [\ion{Fe}{2}] 1.644-\micron, [\ion{O}{1}] 6300 \AA~and [\ion{S}{2}] 6731 \AA~lines only. These low-excitation lines have similar excitation conditions \citep[e.g.,][]{Hollenbach89,Takami02a,Hartigan04b}. We exclude observations of the other lines from this table as a large difference in excitation conditions may cause a systematic difference in the radial velocities \citep{Skinner18,Garufi19,Erkal21b}.

%%%%%%%%%%%%%%%%%%%%%%%%
%%% Table : Radial Velocities
%\begin{rotatetable}

\begin{table*}
\caption{Jet Radial Velocities Measured in the [\ion{Fe}{2}], [\ion{O}{1}],  and [\ion{S}{2}] Lines\label{tbl:vr_lit}}
\begin{tabular}{lcccccc}
\tableline\tableline
Star & Year & Line\tablenotemark{a} & Instrument & Instrument   &$V_\mathrm{rad}$ & Reference\tablenotemark{b} \\
	&	&					&			& Resolution\\
	&	&					&			& (km s$^{-1}$)	& (km s$^{-1}$)
\\ \tableline
RW Aur A	& 2000	& [\ion{S}{2}],[\ion{O}{1}]	& HST-STIS	& 65	& 100 to 140 & 1 \\
		& 2001	& [\ion{Fe}{2}]			& Subaru-IRCS	& 60	& 100 to 140 & 2 \\
		& 2002	& [\ion{O}{1}]			& HST-STIS	& 65	& 100	     & 3 \\
		& {\bf 2012-2021} & {\bf [\ion{Fe}{2}]}	 & Gemini-NIFS	& 55	& {\bf 70 to 130} & {\bf This work} \\
RY Tau	& 2002	& [\ion{O}{1}]			& CFHT-OASIS	 & 135 & --60 & 4 \\
		& 2009	& [\ion{Fe}{2}]			& Gemini-NIFS	& 55	& --70 to --80 & 5 \\
		& {\bf 2012-2021} &  {\bf [\ion{Fe}{2}]} & Gemini-NIFS	& 55	& {\bf --70 to --120} & {\bf This work} \\
DG Tau	& 1998	& [\ion{O}{1}]			& CFHT-OASIS & 90	& --350/--280	& 6 \\ 
		& 1999	& [\ion{S}{2}],[\ion{O}{1}]	& HST-STIS	& 65	& --250 to --350 & 7 \\
		& 2001	& [\ion{Fe}{2}]			& Subaru-IRCS	& 30	& --200 to --250 & 8 \\
		& 2003	& [\ion{O}{1}]			& HST-STIS	& 65	& --180 & 9 \\
		& 2005	& [\ion{Fe}{2}]			& VLT-SINFONI & 100 & --200 & 10 \\
		& 2005--2009 & [\ion{Fe}{2}]		& Gemini-NIFS	& 55	& --170 to --250 & 11 \\
		& {\bf 2013-2019} &  {\bf [\ion{Fe}{2}]} & Gemini-NIFS	& 55	& {\bf --120 to --160} & {\bf This work} \\		
\tableline
\end{tabular}
\tablenotetext{a}{[\ion{Fe}{2}] 1.644 \micron, [\ion{S}{2}] 6731 \AA~and  [\ion{O}{1}] 6300/6363 \AA}
\tablenotetext{b}{(1) \citet{Woitas02,Melnikov09,Liu12}; (2) \citet{Pyo06}; (3) \citet{Coffey04}; (4) \citet{Agra09}; (5) \citet{Coffey15}; (6) \citet{Lavalley00}; (7) \citet{LiuC16}; (8) \citet{Pyo03}; (9) \citet{Coffey07}; (10) \citet{Agra11}; (11) \citet{White14a}}
\end{table*}
%\end{rotatetable}
%%%%%%%%%%%%%%%%%%%%%%%%

For the DG Tau jet, the radial velocities we measured in 2013-2019 are remarkably lower than the previous observations made in 1998--2009 (i.e., --170 to --350 km s$^{-1}$), as for the proper motions listed in Table \ref{tbl:vt_lit}. This trend was also reported by \citet{LiuC16} based on limited epochs of the observations, with the data obtained in 1999 using the {\it Hubble Space Telescope} and ground-based spectro-imaging in 2010 with an angular resolution of $\sim$1\arcsec. The decrease of both the tangential and radial velocities in recent years can be explained if the ejection velocities of the jet knots have decreased, further supporting the moving knot scenario. \citet{LiuC16} pointed out that such a decrease may be related to the expansion of the stellar magnetosphere if the jet is launched from the stellar magnetosphere \citep[or the associated `X-point';][]{Shu00}.

In contrast, the jets from RW Aur A and RY Tau had similar velocities during the period of our observations (2000--2021) and in the past (1930-2010). The tangential velocities tabulated in Tables \ref{tbl:pm:rw} and \ref{tbl:pm:ry} vary between 120 to 210 km s$^{-1}$ and $\sim$120 to 230 km s$^{-1}$, respectively, measured during our observations of 2012-2021. The same physical mechanism that changes the velocity of the jet from DG Tau may also be responsible for these time variations but on shorter timescales.

%%%%%%%%%%%%%%%%%%%%%%%%%%%%%%%%%%%%%%%%
%%% 5.1.2  The Moving-Knot Scenario

\subsubsection{The Shocks Heating/Cooling Scenario} \label{discussion:moving:shocks}

Many authors favor the shock heating scenario for the heating mechanism of the jets close to active pre-main sequence stars. This scenario is in particular favored for DG Tau, for which the jet structures close to the star are spatially resolved. \citet{Lavalley00,Dougados00,Bacciotti00} have resolved bow shock-like structures in the blueshifted jet at 1\arcsec--4\arcsec~from the star.
%, while  \citet{Agra11,White14a} revealed a bubble-like structure in the faint redshifted counterpart.
\citet{Pyo03} observed a distinct low-velocity component ($v \sim -100$ km s$^{-1}$) close to a high-velocity jet knot at 0\farcs8 from the star. The authors interpreted this component as gas entrained by the jet knot. \citet{Lavalley00,Agra11} demonstrated that the observed line intensity ratios at optical and near-infrared wavelengths are also consistent with the shock heating scenario, but more careful analysis may be necessary for gas within 70-100 au \citep[corresponding to 0\farcs4-0\farcs8 for our target stars;][]{Dougados02}. The other analyses that support the shock heating scenario includes \citet{Dougados02,Takami02b,Hartigan04a,Garufi19}. As discussed in Section \ref{results:y}, the structure A in Figure \ref{fig:dg} is similar to an asymmetric bow shock seen in  numerical simulations. 

According to the numerical simulations of shocks by \citet{Hollenbach89}, we would expect a surface brightness of the [\ion{Fe}{2}] emission of $10^{-18}$-$10^{-17}$ W m$^{-2}$ acrsec$^{-2}$ for a shock with a shock velocity of 70-150 km s$^{-1}$ and a pre-shock hydrogen number density of 10$^4$ cm$^{-3}$. In contrast, the peak brightnesses measured for the jet knots reach up to $\sim$$10^{-15}$ W m$^{-2}$ acrsec$^{-2}$, 100-1000 times as large as the modeled values. Such brightnesses, significantly larger than the shock models, could be explained if the hydrogen number density is higher (up to 10$^6$-10$^7$ cm$^{-3}$) or a single knot contains unresolved multiple shock layers. The former explanation is consistent with our measurements of the electron densities (up to $\sim$10$^5$ cm$^{-3}$) if the ionization fraction is low (0.01--0.1) as predicted for the `recombination plateau' in the postshock region, which is primarily responsible for low-ionization forbidden lines such as [\ion{Fe}{2}], [\ion{O}{1}] and [\ion{S}{2}] at $\sim$10$^4$ K \citep[e.g.,][]{Hollenbach89,Hartigan94}.

The different peak intensities between knots (Figures \ref{fig:Ipeak_vs_x} and \ref{fig:Ipeak_vs_JD}; Section \ref{results:I}) could be attributed to different column densities, electron densities, shock velocities, and the filling factors. In Section \ref{results:I} and Figure \ref{fig:Ipeak_vs_JD} we also show different decay timescales for the [\ion{Fe}{2}] intensity. We discuss the implications for this trend in case that the emission is associated with shocks below.

Numerical simulations by \citet{Hartigan94} showed that the optical [\ion{S}{2}] lines, excitation conditions of which are similar to the near-infrared [\ion{Fe}{2}] lines \citep[e.g.][]{Hartigan04b}, have a decay timescale of $\sim$240 days (see Equation \ref{eq:t_decay} for definition) for a shock velocity of 70 km s$^{-1}$, a pre-shock hydrogen number density of 10$^3$ cm$^{-3}$, and an initial magnetic field $B_0$=100 $\mu$G. In practice, the actual shock velocities in our target jets may be significantly larger than 70 km s$^{-1}$, considering the measured jet velocities of 140-270 km s$^{-1}$ (Section \ref{results:v}, after correcting for the jet inclinations). The decay timescales of the emission lines can be even smaller in such conditions, as such shocks yield higher electron densities due to higher temperatures, leading to more rapid cooling \citep{Hartigan94}.
As shown in Figure \ref{fig:t_decay}, the observed decay timescale of the [\ion{Fe}{2}] 1.644-\micron~emission is longer than these values. This would be because, as the shock waves move away, they interact and heat new gas further downstream.

Considering that the jet knot velocities do not change significantly for different epochs (see Tables \ref{tbl:pm:rw}-\ref{tbl:pm:dg}), one would attribute the complicated time variations of the peak intensities in Figure \ref{fig:Ipeak_vs_JD} to different pre-shock conditions. The jet knots appear to spatially expand with time across the jet axis (Figures \ref{fig:FWHM_vs_I}), and also along the jet axis in some cases (Figures \ref{fig:I_x}). One would therefore expect that the gas density in the jet knot would become lower with time, and as a result, the intensity becomes lower as predicted by the shock models \citep{Hollenbach89,Hartigan04b}. Such a trend for electron densities is qualitatively seen in Figure \ref{fig:Ipeak_vs_JD} for some knots.
%Observations with significantly higher angular resolutions and signal-to-noise are required to further investigate a feasibility of this explanation.

Could the shock velocity possibility be significantly smaller than the measured jet velocity? This occurs if the preshock gas is moving forward as well \citep[e.g.,][]{Hartigan87}. \citet{Agra09} estimated a shock velocity of $\sim$20 km s$^{-1}$ based on their observations of the optical [\ion{O}{1}] line and the absence of the optical [\ion{N}{2}] line. However, a slower shock velocity would make the decay timescale for the [\ion{Fe}{2}] emission significantly longer than the observations: using the models for the [\ion{S}{2}] line as above, one would derive  a decay timescale of $\sim$4000 days for a shock with a shock velocity of 35 km s$^{-1}$. Detailed modeling of the [\ion{Fe}{2}] emission for higher electron densities would allow us to further discuss this issue.
 
%%%%%%%%%%%%%%%%%%%%%%%%%%%%%%%%%%%%%%%%
%%% 5.1.3  Other possible heating/cooling mechanisms

\subsubsection{Other Possible Heating/Cooling Mechanisms} \label{discussion:moving:misc}

As described above, the observed trends for the jet knots could be explained with shock heating and cooling. The major reservation of the shock heating and cooling scenario is that we have not been able to spatially resolve the shock structures in the individual knots in our [\ion{Fe}{2}] images close to the star.

The observed chains of the spatially resolved knots may alternatively be due to non-uniform distributions of density or temperature without shocks \citep[e.g.,][]{Shang02,Shang10,Liu12}. This explanation may face the same problem of a short cooling timescale we discussed for shocks described above. In other words, we need a heating mechanism to make the timescale of the intensity decay longer as observed. Such heating could be made through MHD waves \citep{Shang02,Skinner11,Skinner14} or with dissipation of turbulence in the jet \citep{Shang02}. \citet{Shang02} has also discussed heating with ambipolar diffusion, but found that it is more effective in the outer regions of the jet (beyond 500-1000 au, corresponding $>$3\arcsec~from our target stars).

In addition to the above mechanisms, X-ray radiation from the star, or shocks close to the star, may also contribute to the jet heating \citep[e.g.,][]{Shang02,Skinner18}. This heating mechanism should be more efficient closer to the star. It would therefore yield a longer decay timescale for the line intensity closer to the star. However, our observations do not clearly show such a trend (Figure \ref{fig:Ipeak_vs_x}). As the gas densities also affect the cooling timescales, better observations of the electron densities are necessary for further investigating the feasibility of this scenario.

%%%%%%%%%%%%%%%%%%%%%%%%%%%%%%%%%%%%%%%%
%%% 5.1.4  Possible physical link with mass accretion

\subsubsection{Possible Physical Link with Time-Variable Mass Accretion} \label{discussion:moving:acc}

According to the moving-knot scenario, the knots we observed were ejected from the star at irregular intervals for RW Aur A (300--2000 days) and RY Tau (300--1200 days; Section \ref{results:x}). In the case of RW Aur A, \citet{Takami20} revealed a possible time correlation between these knot ejections and optical photometry+spectroscopy, perhaps due to a physical link between jet ejections and mass accretion summarized in Section \ref{intro}. For the jet from DG Tau, the measurements of the time intervals are still tentative, but those for knot C-D and D-E are 1300--1500 days, perhaps far less irregular than those for the other stars.

This possible discrepancy between DG Tau and the other stars may also be related to mass accretion. Both RW Aur A and RY Tau are known to show complicated variabilities in profiles of optical permitted lines, probably signatures of mass accretion \citep{Najita00,Calvet00} even within a few month scale \citep[e.g.,][]{Petrov99,Petrov01a,Alencar05,Facchini16,Takami16,Takami20}. In contrast, \citet{Chou13} made multi-epoch observations for these line profiles for DG Tau in 2010, as well as the other stars, and found that the line profiles observed toward DG Tau were stable during their period of observations. Thorough comparisons between the jet ejections and optical photometry+spectroscopy, like those made by \citet{Takami20} for RW Aur A, are necessary for the other stars as well to further investigate their link.

%%%%%%%%%%%%%%%%%%%%%%%%%%%%%%%%%%%%%%%%
%%%%%%%%%%%%%%%%%%%%%%%%%%%%%%%%%%%%%%%%
%%% 5.2 Stationary shocks?
%%%%%%%%%%%%%%%%%%%%%%%%%%%%%%%%%%%%%%%%
%%%%%%%%%%%%%%%%%%%%%%%%%%%%%%%%%%%%%%%%

\subsection{A Search for Stationary Shocks} \label{discussion:stationary}

The X-ray observations of jets from some protostars and young stars indicate the presence of an inner stationary component in addition to the outer components with proper motions \citep[e.g.,][]{Schneider08,Schneider11}. Such jets include one associated with DG Tau. Multi-epoch observations of the X-ray emission and near-infrared [\ion{Fe}{2}] emission for this star show a stationary component at $\sim$0\farcs2 from the star \citep{Gudel11,White14a}. \citet{Gunther14} conducted model calculations and demonstrated that such shocks can occur due to the recollimation process of the jet near its base.

\begin{figure*}[ht!]
\plotone{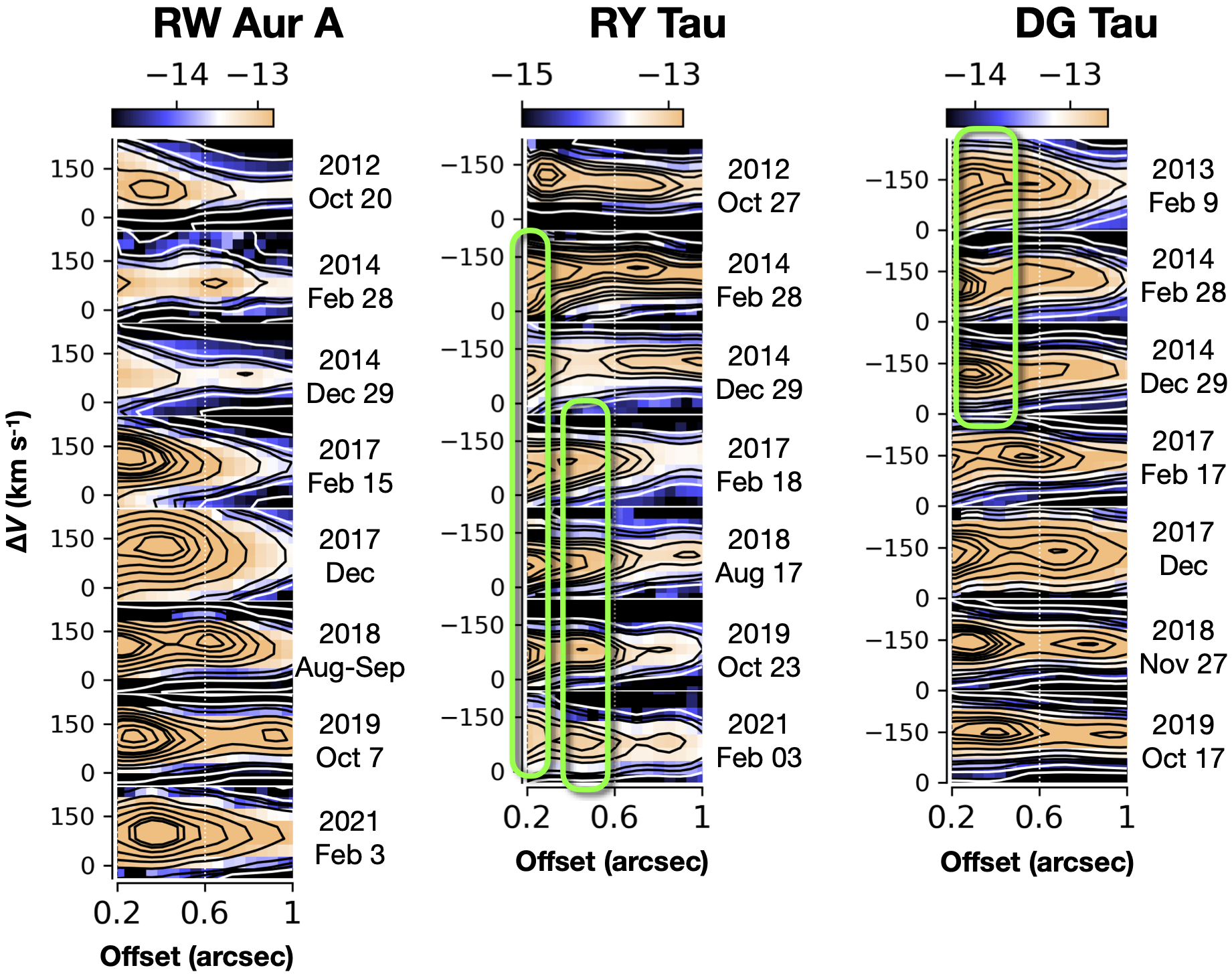}
\caption{Same as  the PV diagrams in Figures \ref{fig:rw}-\ref{fig:dg} but for 0\farcs2 to 1\arcsec from the star and without positional offsets between the epochs. The yellow boxes show possible stationary knots (see text). As for Figures \ref{fig:rw}-\ref{fig:dg}, the contour levels are arbitrarily chosen to show approximate locations of the jet knots. These are identical to Figures \ref{fig:ry} and \ref{fig:dg} for the jets from RY Tau and DG Tau, and slightly revised for RW Aur A for a better display for knots close to the base of the jet.
\label{fig:stationary}}
\end{figure*}

Figure \ref{fig:stationary} shows  the velocity-integrated maps for the base of the three jets without positional offsets. The green boxes show possible stationary shocks. It is difficult to investigate the presence or absence of stationary shocks within $\lesssim$0\farcs2 of the star due to imperfect subtraction of the bright continuum. Observations at better angular resolutions and inner working angles are necessary to confirm or reject this possibility.

%%%%%%%%%%%%%%%%%%%%%%%%
%%% Figure: possible stationary shocks
%%%%%%%%%%%%%%%%%%%%%%%%

%We note that any shock component does not have be observed both in X-ray continuum and the near-infrared [\ion{Fe}{2}] emission due to very different excitation conditions ($T_{\mathrm ex}$$\sim$$10^6$ K and $\sim$$10^4$, respectively). In this context, simultaneous X-ray observations of these jets may also be useful for investigate the nature of the stationary component.

%%%%%%%%%%%%%%%%%%%%%%%%%%%%%%%%%%%%%%%%
%%%%%%%%%%%%%%%%%%%%%%%%%%%%%%%%%%%%%%%%
%%%%%%%%%%%%%%%%%%%%%%%%%%%%%%%%%%%%%%%%
%%% 6. Summary and Conclusions
%%%%%%%%%%%%%%%%%%%%%%%%%%%%%%%%%%%%%%%%
%%%%%%%%%%%%%%%%%%%%%%%%%%%%%%%%%%%%%%%%
%%%%%%%%%%%%%%%%%%%%%%%%%%%%%%%%%%%%%%%%

\section{Summary and Conclusions} \label{summary}

We conducted multi-epoch integral field spectroscopy ($R$=3000-5500, $\Delta v$=55-100 km s$^{-1}$) of near-infrared [\ion{Fe}{2}] emission associated with the well-studied jets from the three active T Tauri stars RW Aur A, RY Tau and DG Tau. The observations were made using Gemini-NIFS, VLT-SINFONI and Keck-OSIRIS with a $\sim$0\farcs1 resolution. During the observations in 2012-2021, we primarily covered the redshifted jet from RW Aur A, and the blueshifted jets from RY Tau and DG Tau, for which we investigate long-term time variabilities in detail. 

Within 3\arcsec~of these stars, we identify a number of knots in the 1.644-\micron~emission, the brightest jet emission line in the spectral coverages of our observations. Most of these, if not all, appear to move outward with proper motions of 0\farcs1-0\farcs4, corresponding to tangential velocities of 70--230 km s$^{-1}$. During our observations, jet ejections from RW Aur A and RY Tau are irregular with time intervals of 300-2000 days. Our data for DG Tau are not sufficient for investigating such a trend, but the measured interval of 1300-1500 day are different from those measured in the past ($\sim$900 and $\sim$1800 days between 1980 and 2005), indicative of an irregularity on a longer timescale.

To investigate potential variabilities of mass accretion or stellar activities over up to $\sim$200 years, we performed comparisons between the measured tangential and radial velocities and those in the literature. For the DG Tau jet, both the tangential ($V_\mathrm{tan}$) and radial velocities ($V_\mathrm{rad}$) seem to have decreased over the past 10--15 years: $V_\mathrm{tan}$ from 100-200 to 70-90 km s$^{-1}$, and $V_\mathrm{rad}$ from 170-350 to 120-160 km s$^{-1}$. In contrast, we do not find any clear evidence for time variation longer than those during our observations over 9 years ($V_\mathrm{tan}$ of 120-210 and 120-240 km s$^{-1}$ for RW Aur A and RY Tau, respectively).

The sizes of the individual knots appear to increase with time across the jet axis, and in some cases along the jet axis as well. In turn, their peak brightnesses in the 1.644-\micron~emission decrease by up to a factor of $\sim$30 during the epochs of our observations. The decay timescale of the emission varies between the knots, and even between different epochs in the same knot, typically ranging between 300 and 3600 days with a median value of $\sim$1000 days. The complexity of their time variations can be explained if the jet knots are unresolved shocks, with the preshock density decreasing toward the downstream but also with some additional spatial variation of the preshock density/temperature.

%Using the 1.533/1.644-\micron~and 1.600/1.644-\micron~intensity ratios, we measured the electron densities of 3$\times$10$^3$ to 1$\times$10$^5$ cm$^{-3}$ at their intensity peaks. These densities are consistent with previous observations of the same jets.

While the overall observed trends for the moving knots are consistent with the shock heating+cooling scenario, our data do not exclude the possibility that those knots are due to non-uniform density/temperature distributions with another heating mechanism such as energy transfer via MHD waves and turbulent dissipation. Furthermore, some of the identified knotty structures may be due to stationary shocks (i.e., without proper motions) associated with the base of the jet. Spatially resolved observations of these knots with significantly higher angular resolutions are necessary to understand their physical nature.

\begin{acknowledgments}
We thank Dr. Elena Valenti for reducing the SINFONI data with the ESO pipeline, and Dr. Lowell Tacconi-Garman for solving the calibration issue with these data sets.
We thank the Gemini Observatory staff for their assistance preparing our programs for data acquisition, and we thank staff observers for executing our program observations during the assigned queue time. 
M.T. is supported by the Ministry of Science and Technology (MoST) of Taiwan (grant No. 106-2119-M-001-026-MY3, 109-2112-M-001-019, 110-2112-M-001-044).
R.G.M. acknowledges support from UNAM-PAPIIT project IN108822 and from CONACyT Ciencia de Frontera project ID 86372.
T.P.R. acknowledges support from the European Research Council through grant No. 743029 (EASY).
This work has made use of data from the European Space Agency (ESA) mission Gaia (https://www.cosmos.esa.int/gaia), processed by the Gaia Data Processing and Analysis Consortium (DPAC, https://www.cosmos.esa.int/web/gaia/dpac/consortium). Funding for the DPAC has been provided by national institutions, in particular the institutions participating in the Gaia Multilateral Agreement.
This research made use of the Simbad database operated at CDS, Strasbourg, France, and the NASA's Astrophysics Data System Abstract Service.
\end{acknowledgments}


\vspace{5mm}
\facilities{Gemini (NIFS), VLT (SINFONI), Keck (OSIRIS)}


\software{
IRAF \citep{Tody86,Tody93},
PyRAF \citep{pyraf},
numpy \citep{numpy},
scipy \citep{scipy},
astropy \citep{astropy}, 
Gemini IRAF package: \citep{Turner06},
ESO Refrex \citep{Fleudling13},
OSIRIS pipeline \citep{Lyke17,Lockhart19},
          }



%\appendix

%\section{Appendix information}


%\bibdata{/Users/hiro/Dropbox/Work/astro.bib}
%\bibliography{/Users/hiro/Dropbox/Work/astro}
\bibliography{v6_rev.bib}

%\bibliography{sample631}{}
\bibliographystyle{aasjournal}



\end{document}